\documentclass[preprint,3p, 12pt,authoryear]{elsarticle}
\usepackage{amssymb}
\usepackage{pdflscape}
\usepackage{multirow}
\usepackage{booktabs}
\usepackage{caption}
\usepackage{subcaption}
\usepackage[edges]{forest}
\usepackage{tabularx}
\usepackage{graphicx}
\usepackage{natbib}
\usepackage{enumitem}

\usepackage{hyperref}

\usepackage[draft]{todonotes}
\usepackage{soul}
\sethlcolor{lime}

\usepackage[framemethod=TikZ]{mdframed}
\usepackage{seqsplit}

\usepackage{float}

\usepackage{adjustbox}

\usepackage{geometry}
\usepackage{tikz}
\usetikzlibrary{shapes, shapes.geometric, matrix, positioning, arrows.meta, fit, calc, backgrounds, shadows, arrows, decorations.pathreplacing}

\usepackage{longtable}

\usepackage{amsmath}

\usepackage{overpic}

\usepackage{setspace}

\definecolor{huberlin-primary}{HTML}{003A6E}
\definecolor{huberlin-secondary}{HTML}{D2C067}
\definecolor{huberlin-lightblue}{HTML}{5b89b0}
\definecolor{huberlin-lightgray}{HTML}{d7dde1}
\definecolor{huberlin-lightsand}{HTML}{fafaf5}
\definecolor{nus-orange}{HTML}{EF7C00}

\journal{European Journal of Operations Research}

\begin{document}
\onehalfspacing

\begin{frontmatter}



\title{Transfer Learning for Loan Recovery Prediction \\ 
under Distribution Shifts with Heterogeneous Feature Spaces\tnoteref{nusstay}}

\tnotetext[nusstay]{Parts of this work were conducted during a research stay of the first author at the National University of Singapore. Access to and use of the Global Credit Data (GCD) dataset were provided through this affiliation.}

\author[inst1]{Christopher Gerling}
\author[inst2]{Hanqiu Peng}
\author[inst2]{Ying Chen}
\author[inst1]{Stefan Lessmann}

\affiliation[inst1]{organization={Humboldt-Universität zu Berlin},%Department and Organization
            addressline={Unter den Linden 6}, 
            city={10099 Berlin},
            country={Germany}}

\affiliation[inst2]{organization={National University of Singapore},
            addressline={10 Lower Kent Ridge Road}, 
            city={Singapore 119076},
            country={Singapore}}
\vspace{-3em}

\begin{abstract}
Accurate forecasting of recovery rates (RR) is central to credit risk management and regulatory capital determination. In many loan portfolios, however, RR modeling is constrained by data scarcity arising from infrequent default events. Transfer learning (TL) offers a promising avenue to mitigate this challenge by exploiting information from related but richer source domains, yet its effectiveness critically depends on the presence and strength of distributional shifts, and on potential heterogeneity between source and target feature spaces.

This paper introduces FT–MDN–Transformer, a mixture-density tabular Transformer architecture specifically designed for TL in RR forecasting across heterogeneous feature sets. The model produces both loan-level point estimates and portfolio-level predictive distributions, thereby supporting a wide range of practical RR forecasting applications. We evaluate the proposed approach in a controlled Monte Carlo simulation that facilitates systematic variation of covariate, conditional, and label shifts, as well as in a real-world transfer setting using the Global Credit Data (GCD) loan dataset as source and a novel bonds dataset as target.

Our results show that FT–MDN–Transformer outperforms baseline models when target-domain data are limited, with particularly pronounced gains under covariate and conditional shifts, while label shift remains challenging. We also observe its probabilistic forecasts to closely track empirical recovery distributions, providing richer information than conventional point-prediction metrics alone. Overall, the findings highlight the potential of distribution-aware TL architectures to improve RR forecasting in data-scarce credit portfolios and offer practical insights for risk managers operating under heterogeneous data environments.

\end{abstract}

% \begin{highlights}
% \todo[inline]{Finalize in the end}
%     \item Propose FT-MDN-Transformer, the first mixture-density tabular Transformer for transfer across heterogeneous feature spaces.  
%     \item Enable recovery-rate (RR) distribution prediction at both the single-loan and portfolio level, directly complementing and extending LGD modeling.  
%     \item Combine Monte Carlo simulations and real-world experiments to assess robustness under covariate, conditional, and label shifts.  
%     \item Demonstrate real-world transfer from a public (Global Credit Data) to a proprietary (UP5) credit dataset with strongly differing feature spaces.  
%     \item Conduct ablations (e.g., categorical encoding, regression vs.\ MDN head) to identify when transfer learning surpasses target-only training.  
% \end{highlights}


\end{frontmatter}



\section{Introduction}
\label{sec:intro-rr}

Recovery rates (RR), defined as the fraction of a loan’s exposure that is recovered after default (i.e.\ $\text{RR} = 1 - \text{LGD}$), play a pivotal role in credit‐risk modeling, provisioning, pricing, portfolio steering, and regulatory capital determination \citep{Bastos2022, Gambetti2019}. Yet robust forecasting of RR is challenging. Recovery outcomes arise from complex, nonlinear interactions among borrower characteristics, loan contract attributes, collateral structure, and macroeconomic conditions \citep{Sopitpongstorn2021}. As a result, the conditional distribution of RR is often irregular and multi‐modal, reflecting, for example, different recovery regimes for secured and unsecured exposures \citep{Gostkowski2020}. This empirical characteristic suggests that focusing exclusively on point estimates may obscure economically and regulatorily relevant information about recovery uncertainty and tail risk.%A further limitation of most existing RR models is that they yield only point estimates, which compress multi-modal structure and may mask distributional risk information even if the predictive $R^2$ seems acceptable.

A second, orthogonal challenge concerns data availability. In banking practice, many portfolios are small, specialized, or niche (for example, shipping finance, project finance, or regionally focused SME lending), so the number of observed defaults, and hence recovery outcomes, is limited. Such data scarcity can impede the training and validation of RR models \citep{Kriebel2020}. A natural response is to leverage information from related portfolios through transfer learning (TL), where a model is pretrained on a richer source portfolio and subsequently adapted to a smaller target portfolio.

However, effective transfer is difficult in RR modeling contexts. First, differences in loan products, collateralization, borrower composition, and macroeconomic conditions can induce substantial discrepancies between source and target portfolios. Statistically, these discrepancies correspond to distributional shifts, including covariate, conditional, and label shifts \citep{Pan2010, Suryanto2022}. Second, portfolios may also differ in the information they recordllateral details, contract terms, or internal classifications may be present in one portfolio but absent from another, resulting in heterogeneous feature spaces with only partial overlap. Most TL approaches assume identical feature schemas across domains and do not directly accommodate heterogeneity; addressing schema mismatch requires ad hoc feature mappings or specialized heterogeneous TL techniques \citep[e.g.,][]{Day2017, Feuz2015, Bellot2019, Sukhija2016}. 

Although TL and distributional recovery modeling address distinct challenges in RR forecasting, they interact in practically relevant ways. In cross‐portfolio settings, reliance on point estimates alone may mask structural differences in recovery behavior and lead to an overly optimistic assessment of transfer success, for example, when average RR transfers well while tail behavior or mixture structure does not. Distributional forecasts, therefore, serve as a complementary tool for validating and governing TL‐based RR models, particularly in contexts where multi-modality and tail risk are central concerns for risk management and regulation.

To address these challenges jointly, we propose \textbf{FT--MDN--Transformer}, a tabular Transformer architecture designed for RR forecasting under data scarcity, heterogeneous feature spaces, and distributional complexity.  The model (i) embeds numeric and categorical features while handling missing or domain-exclusive variables via masking, (ii) supports transfer across non-identical feature spaces through masked tokens and cross-domain fine-tuning, and (iii) outputs full conditional recovery distributions via a mixture density network (MDN) head, enabling both point predictions and portfolio-level density forecasts.
% Proposed revision with 3 contributions
The contributions of this paper are threefold. First, we propose FT–MDN–Transformer, a novel modeling approach that combines tabular Transformers with MDN-based distributional forecasting to enable TL across heterogeneous feature spaces in credit-recovery tasks. 
Second, using two linked real-world recovery datasets (GCD and UP5), we conduct a cross-portfolio empirical study that examines TL performance under realistic constraints, including limited target sample sizes and partial feature overlap.
Third, we develop a Monte Carlo simulation framework with a carefully designed data-generating process for synthetic RR outcomes, enabling controlled analysis under different types and intensities of distributional shift and establishing the boundary conditions and internal validity of TL using our FT–MDN–Transformer.
The comprehensive evaluation of the proposed FT–MDN–Transformer on real and synthetic data yields practical insights into when and how TL can improve RR modeling, providing valuable guidance for model developers, risk managers, and regulators. For research, the simulation framework constitutes a reusable methodological tool for the systematic evaluation of RR forecasting models under specific modeling challenges, including, but not limited to, distributional shift, thereby supporting future work on recovery modeling and related credit risk tasks.

%-----------
% Previous four contribution approach
%The contributions of this paper are fourfold. First, we introduce and link two real-world recovery datasets (GCD and UP5) to form a comprehensive testbed for cross-portfolio RR transfer. Second, we develop a novel modeling approach that combines tabular Transformers with MDN-based distributional forecasting to enable transfer across heterogeneous feature spaces in credit-recovery tasks. Third, we develop a Monte Carlo simulation framework that systematically varies shift regimes and degrees of feature overlap,  and pair it with a real-world GCD→UP5 study to provide the first systematic analysis of TL for RR modeling. Fourth, we distill actionable guidance for risk modelers and regulators by identifying the conditions under which TL is likely to be beneficial, given target sample size, degree of feature overlap, and the magnitude of diagnosed distributional shift.

\section{Related Work}
\label{sec:related-work-rr}

This section reviews three strands of literature that are central to our study: (i) RR and loss given default (LGD) modeling, with emphasis on distributional properties and data scarcity; (ii) TL and domain adaptation in finance; and (iii) recent methods for transfer, adaptation, and robustness in tabular prediction. 
Since RR is the complement of LGD, much of the related work is framed in terms of LGD, reflecting regulatory and accounting conventions. For broader background on deep learning and TL beyond finance, we refer the reader to studies such as \citep{LeCun2015,Goodfellow2016,Weiss2016,Csurka2017}. In the following, we focus on contributions that are directly relevant to cross-portfolio recovery modeling under limited default data, heterogeneous feature spaces, and distribution shifts.

\subsection{Recovery Rate and LGD Modeling}

Recovery rates are core inputs to regulatory and accounting frameworks such as Basel capital requirements, stress testing, and IFRS~9 expected credit loss provisions. Accurate and robust forecasting of recoveries is imperative for risk management and validation. 
%Consequently, a large body of credit-risk research addresses RR modeling to achieve sufficient accuracy and robustness for risk management and validation.
Although regulatory practice is typically expressed in terms of the LGD, recoveries capture the same post-default loss dynamics and serve as the modeling target in this study. 

The credit risk literature emphasizes distinct challenges when modeling RR. A first recurring challenge concerns the \textit{shape} of the RR/LGD distribution. Recoveries are bounded in $[0,1]$ and empirically often exhibit skewness, heavy tails, boundary mass, and multi- or bimodal structures, reflecting, for example, differences between secured and unsecured exposures \citep{Guertler2023}. Early modeling approaches relied on linear or generalized linear regressions, which struggle to capture this heterogeneity \citep{Loterman2012, Calabrese2014}. More flexible specifications have, therefore, been proposed, including mixture models and two-stage approaches that explicitly account for boundary mass and multimodality \citep{Huang2012, Ye2019, Yao2017}. Empirical evidence suggests that no single model class dominates uniformly, with performance depending on the modal structure and tail behavior of the recovery distribution.
A second challenge is that \textit{point forecasts} of the conditional mean are often insufficient for regulatory and risk-management purposes. Capital adequacy, stress testing, and downturn LGD estimation are inherently concerned with adverse scenarios and tail outcomes, which require information about the conditional distribution of recoveries rather than a single expected value. Accordingly, several studies emphasize distributional approaches, including quantile-based methods, mixture specifications, and flexible bounded-response models, to better capture uncertainty and tail risk in recovery outcomes \citep{Gostkowski2020, Sopitpongstorn2021, Gambetti2019}.
Third, \textit{data scarcity} is intrinsic to RR modeling. Recoveries are observed only conditional on default, and defaults are rare in many portfolios, making the estimation of complex models statistically unstable when fitted separately for each book. As a result, most empirical studies focus on single-portfolio settings with proprietary datasets \citep{Ye2019, Bellotti2021, Kriebel2020}. While this abstraction is reasonable for large, well-instrumented portfolios, it leaves open how information can be shared across portfolios when default data are limited and predictor availability differs.

\subsection{Transfer Learning and Domain Adaptation in Finance}

TL studies how predictive models in a target domain can benefit from information learned in a related source domain, rather than being trained exclusively on target data \citep{Zhuang2021}. This perspective directly addresses the data-scarcity problem highlighted above by allowing data-rich portfolios to inform modeling in smaller or less mature books. Differences between source and target distributions, however, pose a central challenge for effective transfer.
Domain adaptation provides a theoretical and methodological framework for addressing such distribution shifts, including covariate, conditional, and label shift \citep{BenDavid2010}. In applied settings, adaptation techniques are typically embedded within broader TL strategies, by, for example, encouraging domain-invariant representations or aligning source and target distributions during training \citep{Ganin2016, Hofer2015}.
Within credit risk, TL and domain adaptation have received increasing attention, but the literature is largely concentrated on probability of default (PD) modeling and binary classification. Early work examines transferring credit-scoring models across related lending domains under limited target labels \citep{Suryanto2019, Suryanto2022}. More recent studies combine TL with explicit adaptation mechanisms to improve robustness under distribution shift \citep{Xu2025, Zhang2025}. Methods that accommodate heterogeneous feature spaces have also been proposed for few-sample credit risk classification \citep{Liu2024}. By contrast, the application of TL to post-default outcomes such as RR and LGD has received substantially less attention than PD modeling. To the best of our knowledge, existing work primarily focuses on PD or binary credit-risk classification, and we are not aware of prior studies that systematically investigate transfer learning for RR/LGD in a distributional modeling framework under heterogeneous feature spaces and controlled distribution shifts.

\subsection{Tabular Transfer, Adaptation, and Robustness Methods}

While much of the TL literature focuses on vision or language, tabular data poses distinct challenges, including heterogeneous feature types, missing values, weak latent structure, and varying feature sets across domains. Recent work, therefore, studies robustness and transfer specifically for tabular prediction under distribution shift. The TableShift benchmark documents substantial performance degradation of tabular models under realistic shifts, even when in-distribution accuracy is strong, and highlights the role of label distribution changes \citep{Gardner2023}.
A related line of research addresses distribution shift via test-time adaptation, adjusting predictions using unlabeled target data. These methods correct for shifted uncertainty, label distributions, or approximately invariant feature relationships, improving robustness without retraining or labeled target data \citep{Kim2025, Zhou2025, Ren2024}. Complementary work highlights the sensitivity of tabular models to feature mismatch across domains, showing that even modest differences in available predictors can sharply degrade performance \citep{Cheng2025}.
In parallel, tabular foundation models have been proposed to improve learning in small-data regimes via large-scale pretraining, achieving strong few-shot and zero-shot performance on diverse benchmarks \citep{Hollmann2025, Gardner2024}. Despite their promise, these approaches are typically evaluated under aligned feature schemas and standard classification or point-regression objectives, leaving open how they apply to distributional recovery modeling across portfolios with partially overlapping features.

\subsection{Gap and Our Positioning}

Taken together, the literature reveals a gap at the intersection of recovery modeling and TL. Existing %RR and LGD 
models are designed to capture complex, often multimodal recovery distributions. However, these models are estimated in single-portfolio settings and assume sufficiently rich, portfolio-specific data. 
This approach is at odds with practice, where recoveries are observed only conditional on rare defaults. When models are transferred or compared across portfolios, additional challenges arise because portfolios differ substantially in size, composition, and available predictors.
At the same time, TL in credit risk has focused primarily on PD modeling and binary classification, while generic tabular transfer and robustness methods are studied under aligned feature spaces and point-prediction objectives. These approaches provide important insights into domain shift and small-sample learning, but they do not directly address distributional recovery modeling across portfolios with heterogeneous and partially overlapping features.
This paper addresses TL for recovery modeling under limited target defaults, heterogeneous feature spaces, and realistic covariate, conditional, and label shifts. Methodologically, we propose a Transformer-based model with a mixture-density output that supports conditional distribution learning and feature-aware transfer. Empirically, we combine real-world recovery datasets with a controlled simulation framework to assess when and how cross-portfolio transfer improves recovery modeling relative to purely in-domain approaches.

\section{Methodology}
\label{sec:methodology-rr}

Our objective is to model recovery rates \(R \in [0,1]\) using an approach that delivers full \emph{distributional} forecasts, supports TL across portfolios with heterogeneous feature schemas, and remains robust under covariate, conditional, and label shifts.  
We achieve this by introducing the \textbf{FT--MDN--Transformer} (\textbf{FT--MDN--T}), a tabular Transformer architecture that combines feature-wise tokenization for heterogeneous data, schema-aware attention masking for partially overlapping features, and a mixture-density head producing conditional RR distributions.

\subsection{FT--MDN--Transformer}
\label{subsec:arch}

\begin{figure}[ht]
    \centering
    \includegraphics[width=0.99\linewidth,trim=2cm 9.5cm 2cm 12cm, clip]{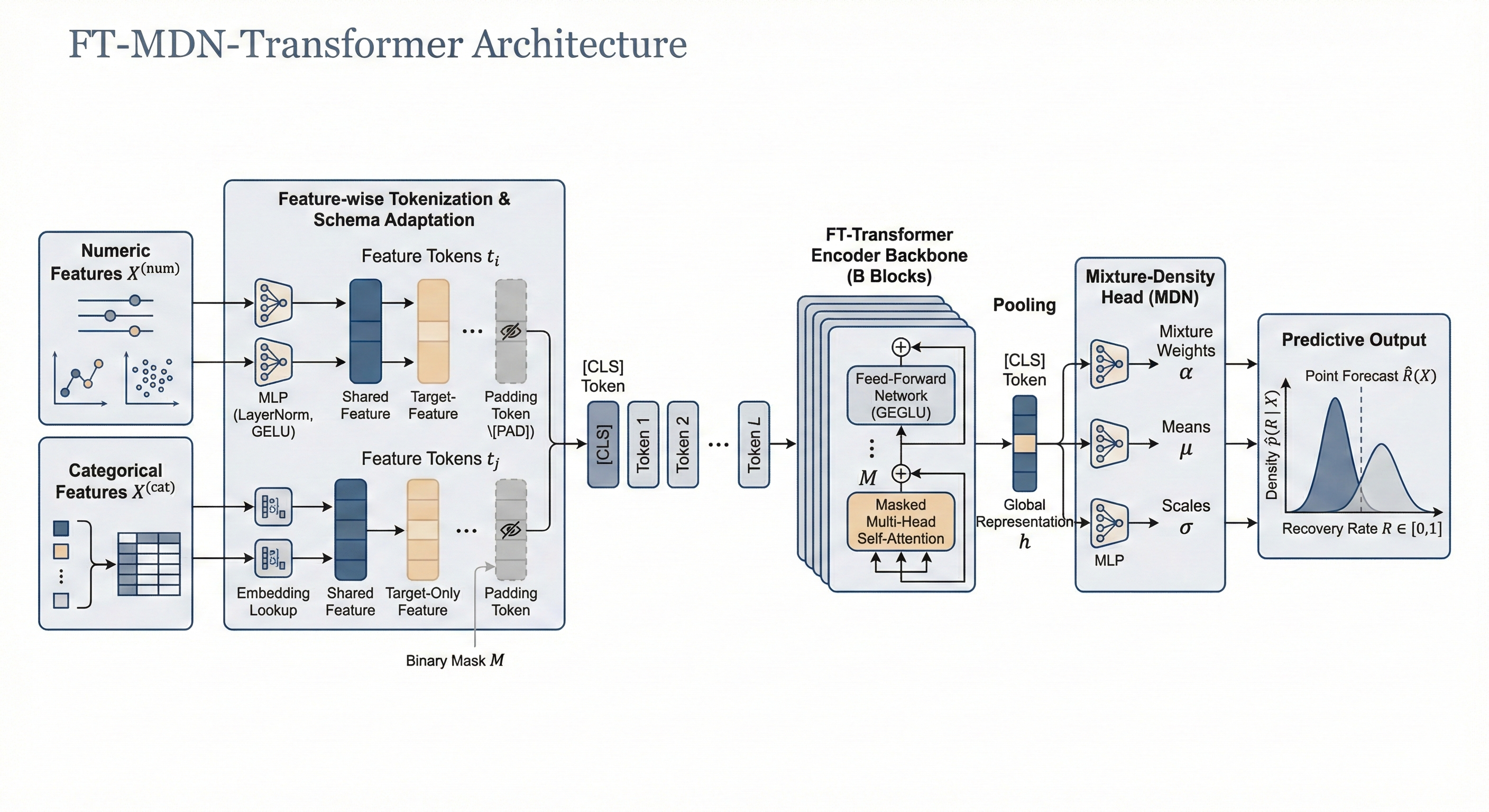}
    \vspace{0.8em}
    \caption[
 FT--MDN--Transformer Architecture
 ]{
 Architecture of the FT--MDN--Transformer with distributional RR predictions.
 }
 \label{fig:ft_mdn_architecture}
\end{figure}

Throughout this section, we use the term \emph{feature} to denote an input variable corresponding to a single column in the tabular schema.
The FT--MDN--T is designed for mixed-type tabular data, where input schemas may differ across domains.  
Its core design principle is feature-wise tokenization: every feature is mapped to an individual token in a fixed-length sequence, enabling the model to activate or deactivate tokens in a controlled, schema-aware manner during transfer.  
A mixture-density head produces full conditional RR distributions, allowing the model to capture multimodality and heteroscedasticity in \(p(R \mid X)\). 
Figure~\ref{fig:ft_mdn_architecture} provides an overview of the model architecture.

\vspace{-2em}

\paragraph{Data structures and notation}

We consider each observation as a vector of \(L\) features, indexed by \(j=1,\dots,L\), where \(L\) denotes the size of the \emph{global} feature universe %considered in a given experiment 
(i.e., the union of source and target features). The feature vector for observation \(i\) is
\vspace{-0.6em}
\[
x_i = (x_{i1}, \dots, x_{iL}),
\]
where features may differ in type (numeric or categorical) and may not be available in all domains.  
A central challenge in the TL setting is that feature schemas are not constant: some features are shared across the source and target data, while others exist only in one domain. Standard neural networks assume a fixed input dimensionality and, therefore, struggle with heterogeneous feature sets. To address this, the FT--MDN--T represents each feature as an individual token, allowing features to be activated or omitted dynamically while preserving a stable overall structure. During pre-training on the source domain, both shared and task-specific features are used to train the backbone; when transferring to a new portfolio, only features that are present in the target schema receive active tokens, while absent features are represented as masked padding tokens.

Each feature is mapped to a \(d\)-dimensional token, yielding the token matrix
\vspace{-0.6em}
\[
T \in \mathbb{R}^{(L+1)\times d},
\vspace{-1em}
\]

where the additional row corresponds to a learnable \texttt{[CLS]} token that pools global information across all features.  
Features that are absent in a domain are filled by a learned padding token and marked in a binary mask
\vspace{-1em}
\[
M \in \{0,1\}^{L+1},
\vspace{-0.6em}
\]

which prevents attention from flowing into or out of these positions.  
The Transformer encoder processes \((T, M)\), producing contextualized feature embeddings
\vspace{-0.6em}
\[
Z \in \mathbb{R}^{(L+1)\times d},
\vspace{-1em}
\]

and the pooled representation \(h = Z_{\texttt{[CLS]}}\) summarizes all available feature information for downstream prediction.

\paragraph{Feature-wise tokenization}
Feature-wise tokenization provides a flexible, expressive interface between raw tabular data and the Transformer architecture.  
Instead of concatenating all features into a single vector---as done in conventional tabular models---we encode each feature independently and preserve this separation throughout the network.  
This separation is crucial for TL: it allows the model to retain shared features while cleanly masking unavailable ones, without altering the geometry of the input space.

Lightweight MLPs transform numeric features,
\vspace{-0.6em}
\[
g_j : \mathbb{R} \to \mathbb{R}^{d}, \qquad t_{ij} = g_j(x_{ij}),
\vspace{-0.6em}
\]
which enable nonlinear rescaling and feature-specific preprocessing.  
Categorical features are embedded using learned lookup tables,
\vspace{-0.6em}
\[
E_j \in \mathbb{R}^{K_j \times d}, \qquad t_{ij} = E_j[x_{ij}],
\vspace{-0.6em}
\]
where \(K_j\) is defined over the union of \emph{source-train} and \emph{target-train} categories for feature \(j\) (constructed without using validation/test data).  
Categories that occur only in the target training split, therefore, have dedicated embedding rows already at transfer time; these rows are unused (and hence not updated) during pretraining and are optimized once the model is fine-tuned on target data.  
A dedicated \texttt{<UNK>} entry is reserved exclusively for categories that appear only at validation or inference time (i.e., not observed in either training split), ensuring robust prediction without information leakage.
Because TL requires the ability to expand or reduce the available feature set, the model maintains separate encoder banks for shared and task-specific features.  
Task-specific features that do not exist in the current domain are replaced by a learned padding token and masked in attention, ensuring that the model neither relies on nor is distorted by non-existent features, while still benefiting from their contribution during pre-training in domains where they are observed.

\paragraph{Transformer backbone}
Once tokenized, the feature representations are processed by a stack of \(B\) encoder blocks following the FT-Transformer architecture \citep{Gorishniy_TF-Transformer}.  
The backbone enables the model to learn rich interactions between features---both linear and nonlinear ---through multi-head self-attention.  
This is particularly valuable for credit recovery modelling, where dependencies among borrower characteristics, loan terms, and collateral indicators can be highly structured and predictive.

Each encoder block applies pre-normalized self-attention with the mask \(M\), ensuring that features absent in the current domain are ignored while all available shared features remain fully usable.  
A GEGLU-activated feed-forward network further transforms the representations, and residual connections stabilize optimization throughout the stack.

As the feature tokens propagate through the encoder, each block incrementally refines them by integrating information from all other available features via masked self-attention.  
After the final block, the model produces the contextualized token matrix \(Z\), where every token encodes both its own feature value and the interactions it has learned with the rest of the feature set.

The contextualized \texttt{[CLS]} token \(h\) serves as a compact, permutation-aware summary of the full feature set and their interactions, and is used to parameterize the predictive distribution in the output head.

\paragraph{Mixture-density head}
Recovery rates exhibit bounded support, multimodality, and heteroscedasticity\allowbreak---properties that standard regression heads cannot represent.  
To address this, the FT--MDN--Transformer employs a mixture-density network head, which maps the pooled representation \(h\) to the parameters of a mixture distribution:
\vspace{-0.6em}
\[
\hat p(R \mid X)
= \sum_{k=1}^{K} \alpha_k \, \mathcal{N}(R \mid \mu_k, \sigma_k^2).
\vspace{-0.6em}
\]

The MDN naturally captures multiple recovery scenarios (e.g., near-zero recovery vs. partial recovery) and provides a full predictive distribution rather than a single point estimate.  
Valid parameters are obtained through
\vspace{-1em}
\[
\alpha = \mathrm{softmax}(z_{\alpha}/\tau), \qquad
\mu = \mathrm{sigmoid}(z_{\mu}) \cdot \mu_{\max}, \qquad
\sigma = \mathrm{softplus}(z_{\sigma}) + \varepsilon,
\vspace{-1em}
\]
ensuring positivity and alignment with the bounded support of recovery rates.

Training proceeds by minimizing the negative log-likelihood of observed recovery values.  
For practitioners requiring point forecasts, we report the mixture mean,
\vspace{-0.6em}
\[
\hat{R}(X) = \sum_{k=1}^{K} \alpha_k \mu_k,
\vspace{-0.6em}
\]
while retaining access to the full predictive distribution for downstream portfolio applications.  
A simplified ablation variant (\textit{FT--Reg}) replaces the MDN with a linear head to benchmark the value of distributional modeling.

\vspace{0.25em}
\subsection{Heterogeneous Schema Adaptation}
\label{subsec:hetero}

In cross-portfolio transfer, source and target portfolios often differ in feature coverage.  
Let \(\mathcal{F}_{\mathrm{s}}\) and \(\mathcal{F}_{\mathrm{t}}\) denote the sets of available source and target features, respectively;  
their intersection \(\mathcal{F}_{\cap}=\mathcal{F}_{\mathrm{s}}\cap\mathcal{F}_{\mathrm{t}}\) are the shared features,  
\(\mathcal{F}_{\mathrm{s}\setminus\mathrm{t}}\) are features present only in source, and  
\(\mathcal{F}_{\mathrm{t}\setminus\mathrm{s}}\) are features present only in target.

\paragraph{Token-level mechanism}
FT--MDN--T resolves schema mismatch entirely at the token level.  
Each feature corresponds to a fixed position in the token sequence, so differences between source and target schemas are handled by reusing, omitting, or adding tokens---without changing the backbone.

\begin{itemize}
\item \textbf{Shared features} (\(\mathcal{F}_{\cap}\)):  
      Pretrained tokenizers and embeddings are reused for all features that appear in both source and target domains (including categorical vocabularies defined over the union of source-train and target-train categories).  
      During fine-tuning, these parameters can be frozen initially and later unfrozen for joint adaptation.  
\item \textbf{Features present only in the source at transfer} (\(\mathcal{F}_{\mathrm{s}\setminus\mathrm{t}}\)):  
      The model is reconfigured to the target schema, excluding these features from fine-tuning.  
      Knowledge learned from them during pretraining remains encoded in the shared representations and backbone weights, maintaining continuity across domains.  
      Optionally, random feature masking during pretraining can simulate future feature absence, helping the backbone generalize to partially observed schemas.
\item \textbf{Features present only in the target at transfer} (\(\mathcal{F}_{\mathrm{t}\setminus\mathrm{s}}\)):  
      New tokenizers and embeddings are initialized for these features and trained on the target data, while the pretrained backbone and head are reused and jointly updated.

\end{itemize}

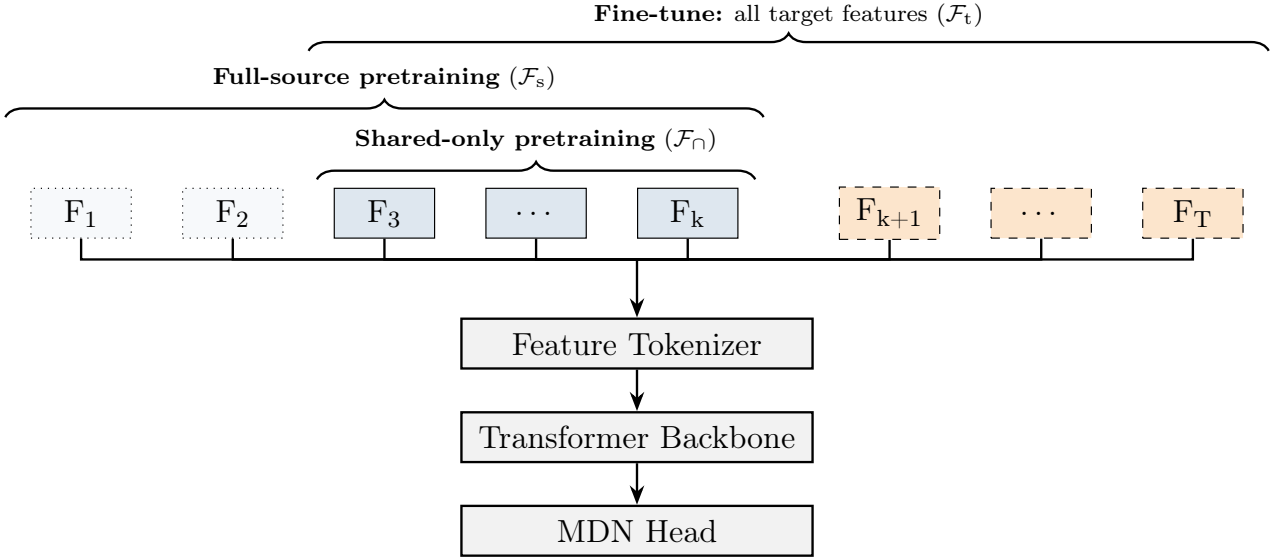
\begin{figure}[ht]
\centering
\scalebox{1.1}{
\begin{tikzpicture}[
    font=\small,
    srconly/.style={rectangle, draw, minimum width=1.2cm, minimum height=0.6cm, fill=huberlin-lightblue!5, dotted},
    shared/.style={rectangle, draw, minimum width=1.2cm, minimum height=0.6cm, fill=huberlin-lightblue!20},
    newemb/.style={rectangle, draw, minimum width=1.2cm, minimum height=0.6cm, fill=nus-orange!20, dashed},
    model/.style={rectangle, draw, thick, minimum width=4.2cm, minimum height=0.6cm, fill=gray!10},
    arrow/.style={->, thick, >=Stealth},
    brace/.style={decorate, decoration={brace, amplitude=6pt}, thick},
    node distance=0.1cm and 0.6cm
]
\node[srconly] (s1) at (-1.8,0) {F\textsubscript{1}};
\node[srconly, right=of s1] (s2) {F\textsubscript{2}};
\node[shared, right=of s2] (f1) {F\textsubscript{3}};
\node[shared, right=of f1] (f2) {\dots};
\node[shared, right=of f2] (fs) {F\textsubscript{k}};
\node[newemb, right=1.2cm of fs] (ft1) {F\textsubscript{k+1}};
\node[newemb, right=of ft1] (ft2) {\dots};
\node[newemb, right=of ft2] (ftt) {F\textsubscript{T}};

\node[coordinate] (mergeShared) at ($(s1.south)!0.5!(ftt.south)-(0,0.25)$) {};
\node[model, below=0.7cm of mergeShared] (tokenizer)   {Feature Tokenizer};
\node[model, below=0.5cm of tokenizer]   (transformer) {Transformer Backbone};
\node[model, below=0.5cm of transformer] (mdn)         {MDN Head};

\foreach \x in {s1,s2,f1,f2,fs,ft1,ft2,ftt} { \draw[thick] (\x.south) |- (mergeShared); }
\draw[arrow] (mergeShared) -- (tokenizer.north);
\draw[arrow] (tokenizer.south) -- (transformer.north);
\draw[arrow] (transformer.south) -- (mdn.north);

\draw[brace, draw=black]  ([xshift=-0.3cm,yshift=0.85cm]s1.north west) -- ([xshift=0.3cm,yshift=0.85cm]fs.north east)
  node[midway, above=5pt, font=\scriptsize]{\textbf{Full-source pretraining} (\(\mathcal{F}_{\mathrm{s}}\))};
\draw[brace, draw=black]  ([xshift=-0.2cm,yshift=0.1cm]f1.north west) -- ([xshift=0.2cm,yshift=0.1cm]fs.north east)
  node[midway, above=5pt, font=\scriptsize]{\textbf{Shared-only pretraining} (\(\mathcal{F}_{\cap}\))};
\draw[brace, draw=black]  ([xshift=-0.3cm,yshift=1.65cm]f1.north west) -- ([xshift=0.3cm,yshift=1.65cm]ftt.north east)
  node[midway, above=4pt, font=\scriptsize]{\textbf{Fine-tune:} all target features (\(\mathcal{F}_{\mathrm{t}}\))};
\end{tikzpicture}
}
\vspace{0.3em}
\caption[
Two-Stage Schema Adaptation Strategies
]{
Two-stage schema adaptation comparing shared-only and full-source pretraining followed by expansion to the target schema.
}
\label{fig:ft_mdn_training_heterogeneous}
\end{figure}

\paragraph{Two-stage transfer schedule}
We train FT--MDN--T in a two-stage transfer setup comprising pretraining on a labeled source domain and fine-tuning on a target domain with partially overlapping features.  
During pretraining, the model learns generalizable feature interactions on the source schema \(\mathcal{F}_{\mathrm{s}}\) using AdamW (optionally with cosine-annealed learning rates) and early stopping on a held-out source validation split.  
We explore two complementary strategies, illustrated in Figure~\ref{fig:ft_mdn_training_heterogeneous}.  
In the first, \emph{shared-only} pretraining, we restrict training to the intersection \(\mathcal{F}_{\cap}=\mathcal{F}_{\mathrm{s}}\cap\mathcal{F}_{\mathrm{t}}\), ensuring full feature alignment between domains at the cost of discarding non-shared source information.  
In the second, \emph{full-source} pretraining, the model uses the entire source schema \(\mathcal{F}_{\mathrm{s}}\); at transfer, tokens corresponding to features missing in the target (\(\mathcal{F}_{\mathrm{s}\setminus\mathrm{t}}\)) are replaced by a learned \texttt{PAD} vector and masked from attention. In contrast, tokens for new target-only features (\(\mathcal{F}_{\mathrm{t}\setminus\mathrm{s}}\)) are instantiated with freshly initialized tokenizers and embeddings.  
Because the Transformer operates on a fixed feature order, both strategies preserve the positional structure learned during pretraining and enable seamless schema reuse without manual feature mapping.

During the \emph{fine-tuning} stage on the target dataset, pretrained parameters serve as initialization, and the learning rate is reduced to stabilize adaptation.  
In the first \(e_{\mathrm{warm}}\) epochs, shared-token embeddings and early Transformer blocks can be frozen to maintain previously learned representations; afterwards, all modules are unfrozen for joint optimization.  
New target-only features are trained from scratch while remaining components continue to update via gradient flow, allowing the model to incorporate additional feature information without losing previously learned structure.  
This two-stage process provides a coherent mechanism for reusing pretrained weights, expanding schemas dynamically, and achieving stable convergence even when \(n_{\mathrm{t}}\ll n_{\mathrm{s}}\).

\vspace{0.5em}
\subsection{Experimental Setup and Evaluation}
\label{subsec:exp_setup}

We now describe the experimental protocol used consistently across both the real-data and simulation studies, covering preprocessing, training regimes, and evaluation metrics.

\paragraph{Preprocessing}
Numerical features are standardized using \emph{source-train} statistics, and the same scalers are applied during fine-tuning and evaluation to mimic deployment.  
Categorical variables are kept as they occur in practice and represented through integer indices with trainable embeddings.  
For each categorical feature, the embedding vocabulary is constructed once from the union of \emph{source-train} and \emph{target-train} categories (excluding validation/test data), so categories first observed in the target domain already have dedicated embedding rows at transfer time (initialized but not updated during pretraining, and learned during fine-tuning).  
A dedicated \texttt{<UNK>} token is reserved exclusively for categories that occur only in the validation or test data and were not observed in either training split.  
Dummy encoding is applied only in ablation studies and for baseline models (e.g., tree ensembles) that cannot natively handle embeddings.  
Feature order is fixed across all experiments to preserve consistent token positions, ensuring that pretraining and fine-tuning operate on aligned feature indices and positional embeddings.  
All experiments use stratified train--validation splits with 20\% validation share and identical random seeds across models for comparability.

\begin{figure}[ht]
\centering
\scalebox{1.1}{
\begin{tikzpicture}[
    node distance=0.5cm, auto, font=\footnotesize,
    classBlock/.style={rectangle, draw, fill=huberlin-lightsand!60, text width=4em, align=center, rounded corners, minimum height=3em},
    sourceBlock/.style={rectangle, draw, fill=huberlin-lightblue!30, text width=4em, align=center, rounded corners, minimum height=3em},
    targetBlock/.style={rectangle, draw, fill=nus-orange!30, text width=4em, align=center, rounded corners, minimum height=3em},
    arrow/.style={thick, ->, >=stealth}
]
\node[font=\bfseries] at (-1.3,3.5) {Zero-Shot};
\node[font=\bfseries] at (3.8,3.5)  {Target-Baseline};
\node[font=\bfseries] at (9.15,3.5) {Transfer-Learning};
\draw[dashed, gray] (1.25,3.7) -- (1.25,-2.8);
\draw[dashed, gray] (6.55,3.7) -- (6.55,-2.8);

\node[classBlock] (S1Model) at (0,2)  {Model\\Training};
\node[classBlock] (S1Eval)  at (0,0)  {Evaluation};
\node[sourceBlock, left=0.8cm of S1Model] (S1Source) {Source};
\node[targetBlock, left=0.8cm of S1Eval]  (S1Target) {Target};
\draw[arrow] (S1Source) -- (S1Model);
\draw[arrow] (S1Model) -- (S1Eval);
\draw[arrow] (S1Target) -- (S1Eval);

\node[classBlock] (S2Model) at (5.3,2) {Model\\Training};
\node[classBlock] (S2Eval)  at (5.3,0) {Evaluation};
\node[targetBlock, left=0.8cm of S2Model] (S2TargetTrain) {Target};
\node[targetBlock, left=0.8cm of S2Eval]  (S2TargetEval) {Target};
\draw[arrow] (S2TargetTrain) -- (S2Model);
\draw[arrow] (S2Model) -- (S2Eval);
\draw[arrow] (S2TargetEval) -- (S2Eval);

\node[classBlock] (S3Pre)   at (10.6,2) {Model\\Pretrain};
\node[classBlock] (S3Fine)  at (10.6,0) {Model\\Fine-tune};
\node[classBlock] (S3Eval)  at (10.6,-2){Evaluation};
\node[sourceBlock, left=0.8cm of S3Pre]  (S3Source) {Source};
\node[targetBlock, left=0.8cm of S3Fine] (S3TargetTrain) {Target};
\node[targetBlock, left=0.8cm of S3Eval] (S3TargetEval) {Target};
\draw[arrow] (S3Source) -- (S3Pre);
\draw[arrow] (S3Pre) -- (S3Fine);
\draw[arrow] (S3TargetTrain) -- (S3Fine);
\draw[arrow] (S3Fine) -- (S3Eval);
\draw[arrow] (S3TargetEval) -- (S3Eval);
\end{tikzpicture}
}
\vspace{0.4em}
\caption[
Evaluation Scenarios
]{
Evaluation scenarios used throughout, including zero-shot, target baseline,
and transfer learning.
}
\label{fig:evaluation_scenarios}
\end{figure}
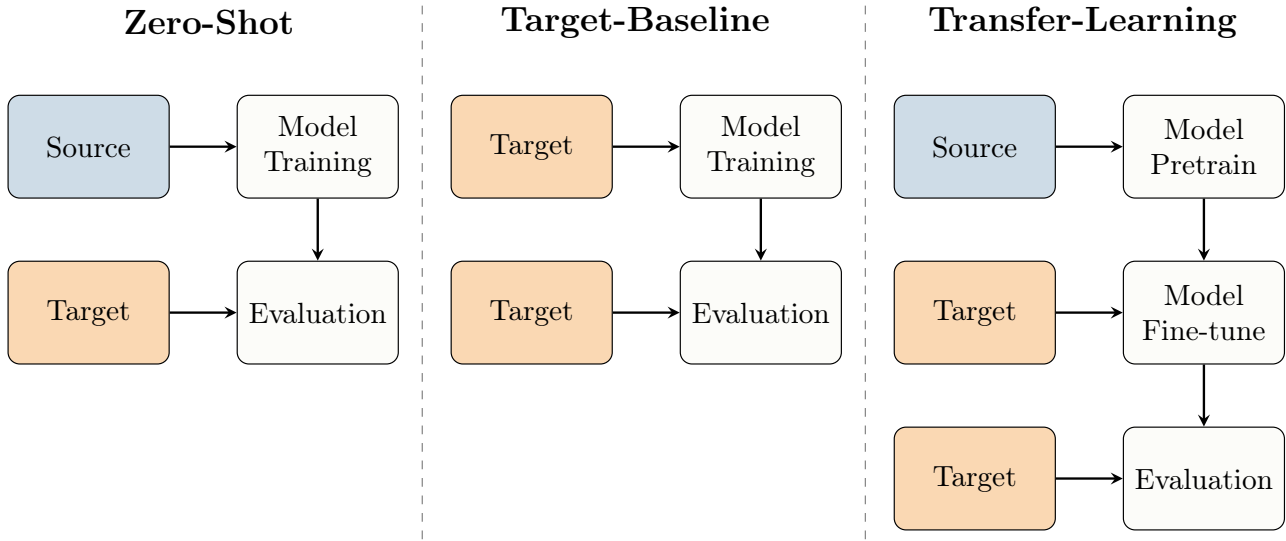

\paragraph{Evaluation scenarios and metrics}
We assess model performance under three complementary training regimes with aligned data splits and identical optimization settings:  
(1) Zero-Shot, where models are trained on the source and evaluated directly on the target without adaptation;  
(2) Target-Baseline, where models are trained from scratch using only target data; and  
(3) Transfer-Learning, where models are first pretrained on the source and then fine-tuned on the target following the two-stage schedule.  
These three scenarios are summarized in Figure~\ref{fig:evaluation_scenarios}.

For quantitative assessment, we report the coefficient of determination (\(R^2\)) and mean absolute error (MAE) on held-out validation sets as the main accuracy metrics.  
In addition, because FT--MDN--T produces conditional density estimates, we visualize predicted and empirical RR distributions to assess calibration and multimodality.  
At the portfolio level, individual loan densities
\(\hat p_i(R)=\sum_{k=1}^{K}\alpha_{ik}\,\mathcal{N}(R\mid \mu_{ik},\sigma_{ik}^2)\)
are averaged into a mixture-of-mixtures
\(\bar p(R)=\tfrac{1}{m}\sum_{i=1}^{m}\hat p_i(R)\),
which is compared with the empirical recovery histogram to analyze overall shape, peaks, and tail behavior.  
These plots serve as qualitative diagnostics complementing the quantitative accuracy metrics.

\vspace{0.5em}

\vspace{0.5em}
\subsection{Baselines and Ablations}
\label{subsec:ablations}

We benchmark FT--MDN--T against established nonlinear tabular models, specifically XGBoost, Random Forest, a multilayer perceptron (MLP), and an FT--Transformer variant with a deterministic regression head (FT--Reg). All models rely on the same data splits, preprocessing steps, and early-stopping procedure to ensure a consistent and fair comparison. The ablation analysis focuses on three aspects of the framework: the effect of using a probabilistic MDN head rather than a deterministic regression head, the influence of pretraining on either the shared subset of features or the full source schema, and the role of different categorical encoding strategies.

Further baseline configurations, training settings, and hyperparameters are documented in ~\ref{app:impl}. The complete reproducibility package, including configuration files and scripts, is available in the accompanying code repository.\footnote{\url{https://github.com/christopher3996/recovery_rates_transfer_learning}}

\section{Real-Data Study}
\label{sec:recovery-data}

This section reports the real-data experiments that provide the empirical setting for assessing cross-domain transfer in RR modeling. We first introduce the GCD (source) and UP5 (target) datasets and report descriptive statistics, including bimodality and feature overlap. Despite hundreds of candidate features, only 37 overlap, illustrating the scope of schema mismatches encountered in practice. At the same time, this particular source--target pairing represents a particularly challenging transfer scenario, as the limited overlap and pronounced portfolio differences place strong demands on any TL approach.

Following the introduction of the data, we clarify our experimental design by organizing the empirical analysis into four increasingly challenging steps. These steps move from in-domain distributional modeling to representation robustness, heterogeneous-schema expansion, and finally cross-portfolio transfer. For each step, we report and interpret the corresponding results, thereby providing a structured empirical assessment of the proposed framework under increasing levels of heterogeneity and distributional mismatch.

\subsection{Real-Data Description (GCD/UP5)}

We use the \emph{Global Credit Data} (GCD) consortium database as the \textbf{source domain} and the \emph{UP5} portfolio as the \textbf{target domain}. Both datasets consist of defaulted corporate exposures, but differ substantially in product type, feature coverage, and recovery behavior.

GCD is a consortium dataset comprising data from 55+ banks. It supports key credit risk modeling parameters, including LGD, PD, and Exposure at Default. In this study, we use the LGD database from GCD, which contains detailed information on loan attributes, collateral, borrowers, guarantors, and loan histories collected at the time of default. Recovery rates are the forecasting target. They are computed as the present value of net cumulative recoveries divided by default exposure if the unresolved default duration exceeds 2{,}520 days, or determined via a lookup table otherwise. Table~\ref{tab:cohort_overview} reports raw counts/features alongside the final modeling samples used in the empirical analysis.

\begin{table}[ht!]
\centering
\small
\caption[
Core Sample Characteristics of the Two Domains
]{
Core sample characteristics of the two domains (raw and final after cleaning).
}
\label{tab:cohort_overview}
\begin{tabularx}{\textwidth}{>{\raggedright\arraybackslash}p{7.5cm} p{4.5cm}p{4cm}}
\toprule
 & \textbf{GCD (Source)} & \textbf{UP5 (Target)} \\
\midrule
Observation window & 1983--2023 & 1996--2023 \\
Sample size ($N$; raw / final) & 331{,}887 / 14{,}050 & 2{,}170 / 1{,}725 \\
Instrument type & Secured loans & Unsecured bonds \\
Observation unit & Loan facility & Bond issuance \\
Geographic coverage & US and Europe & Predominantly US \\
Features (pre-dummy; raw / final) & 147 / 73 & 256 / 164 \\
Numeric / categorical (final) & 50 / 23 & 155 / 9 \\
\bottomrule
\end{tabularx}
\end{table}
\vspace{1em}

Originally, the GCD database for LGD contained 331{,}887 defaulted loan facilities with 147 features, spanning defaults from April~1983 to August~2023. It includes firm-specific and security-specific characteristics as well as 36 macroeconomic variables. For forecasting purposes, we first removed variables that are purely identifiers, fields used only for the computation of recovery rates or currency conversions, and date fields after transforming them into time-to-default or macro-alignment features.

Within the remaining set of original and derived variables, 58 features contained missing values, of which 37 had more than 50\% missing observations, and 21 had more than 90\%. To avoid heavy reliance on imputation, we discard all features with more than 50\% missing values and then drop any observation that is still incomplete on this reduced feature set. This leaves 73 features (including the 36 macroeconomic variables) and 32{,}563 fully observed loan facilities, i.e.\ roughly 10\% of the original 331{,}887 records.

Several key categorical variables (e.g.,\ internal risk rating, seniority code, facility type, industry, and country-of-business indicators) also contain an explicit ``unknown'' category. To ensure economically interpretable covariates, we remove all observations for which any of these features is coded as unknown, reducing the sample to 14{,}306 facilities. Finally, we trim extreme RR outliers to the economically plausible range $[0, 1.2]$, resulting in a final GCD working sample of 14{,}050 observations used in the empirical analysis.

The UP5 dataset, developed under the National Research Foundation project at the National University of Singapore, contains 1{,}725 observations from 576 firms over 1996--2023. It includes 256 candidate features gathered from Refinitiv and Bloomberg (macroeconomic, firm-level, and bond-level), as well as additional credit analytics variables from the NUS Credit Research Initiative. After preprocessing, the final UP5 feature set contains 164 features and 1{,}725 observations (Table~\ref{tab:cohort_overview}). RR is defined as the average ratio of a bond's dirty price to its par value within 30 days post-default.

%--------------------------------------------------------------------
\paragraph{Descriptive evidence}\label{par:descriptive-evidence}

\begin{figure}[ht!]
  \centering
  \begin{minipage}{0.3\textwidth}
    \centering
    \textbf{GCD Recovery Rates} \\
    \includegraphics[width=\linewidth]{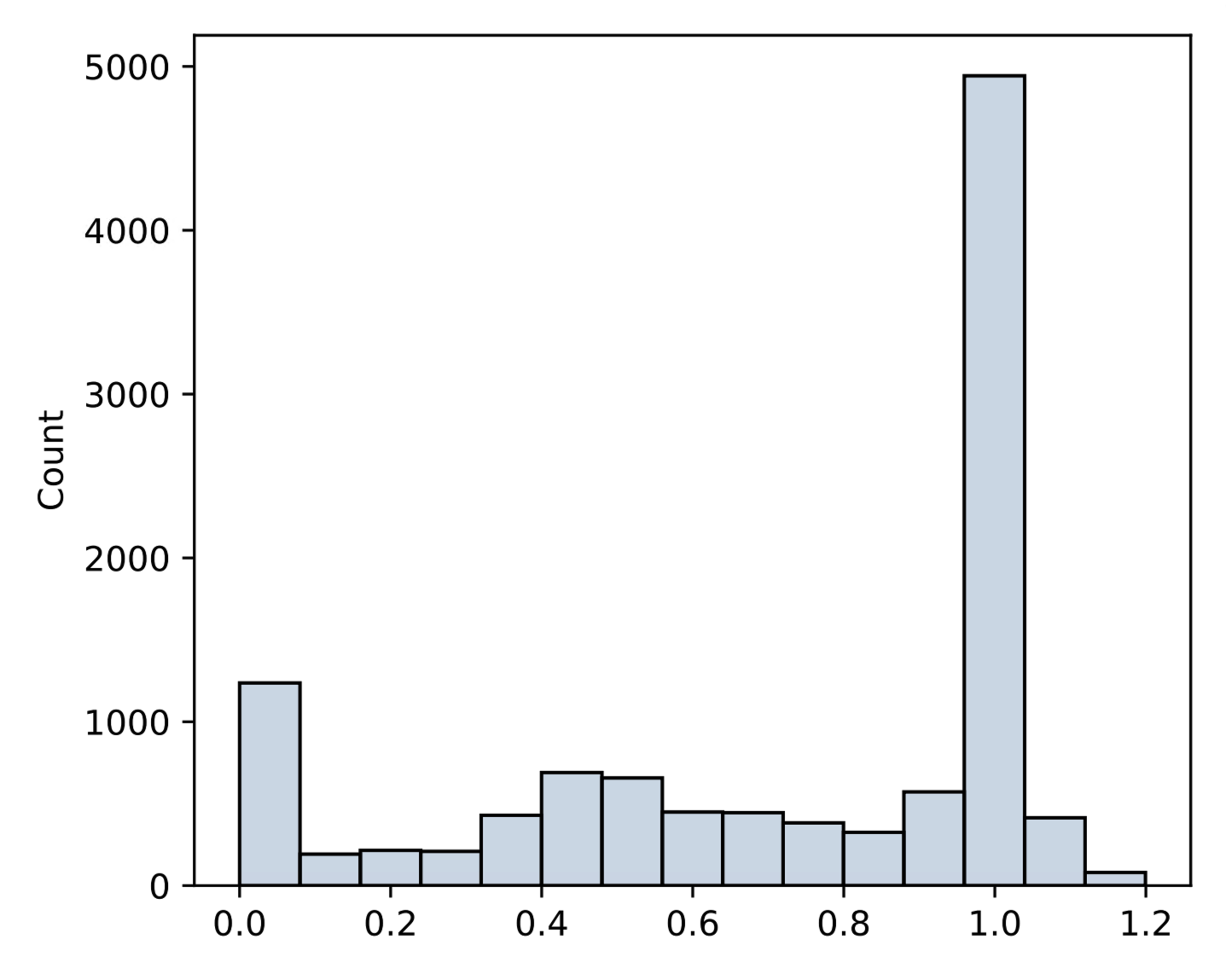} \\
  \end{minipage}
  \hfill
   \raisebox{0.08cm}{
  \begin{minipage}{0.35\textwidth}
    \centering
    \textbf{Feature Overlap} \\
   \begin{tikzpicture}[font=\small]
  \begin{scope}[yshift=-0.15cm]
    \filldraw[fill=huberlin-lightblue!30, draw=huberlin-lightblue!60, fill opacity=0.4]
      (-0.8,0) circle (1.5cm);
    \filldraw[fill=nus-orange!30, draw=nus-orange!60, fill opacity=0.4]
      (0.8,0) circle (1.8cm);
    \node[font=\bfseries] at (-0.1,0) {37};
    \node[font=\itshape, gray] at (-1.6,0) {36};
    \node[font=\itshape, gray] at (1.5,0) {127};
  \end{scope}
  \node[above] at (-1.1,1.6) {GCD (73)};
  \node[above] at (1.1,1.6) {UP5 (164)};
\end{tikzpicture}
  \end{minipage}
  }
  \hfill
  \begin{minipage}{0.3\textwidth}
    \centering
    \textbf{UP5 Recovery Rates} \\
    \includegraphics[width=\linewidth]{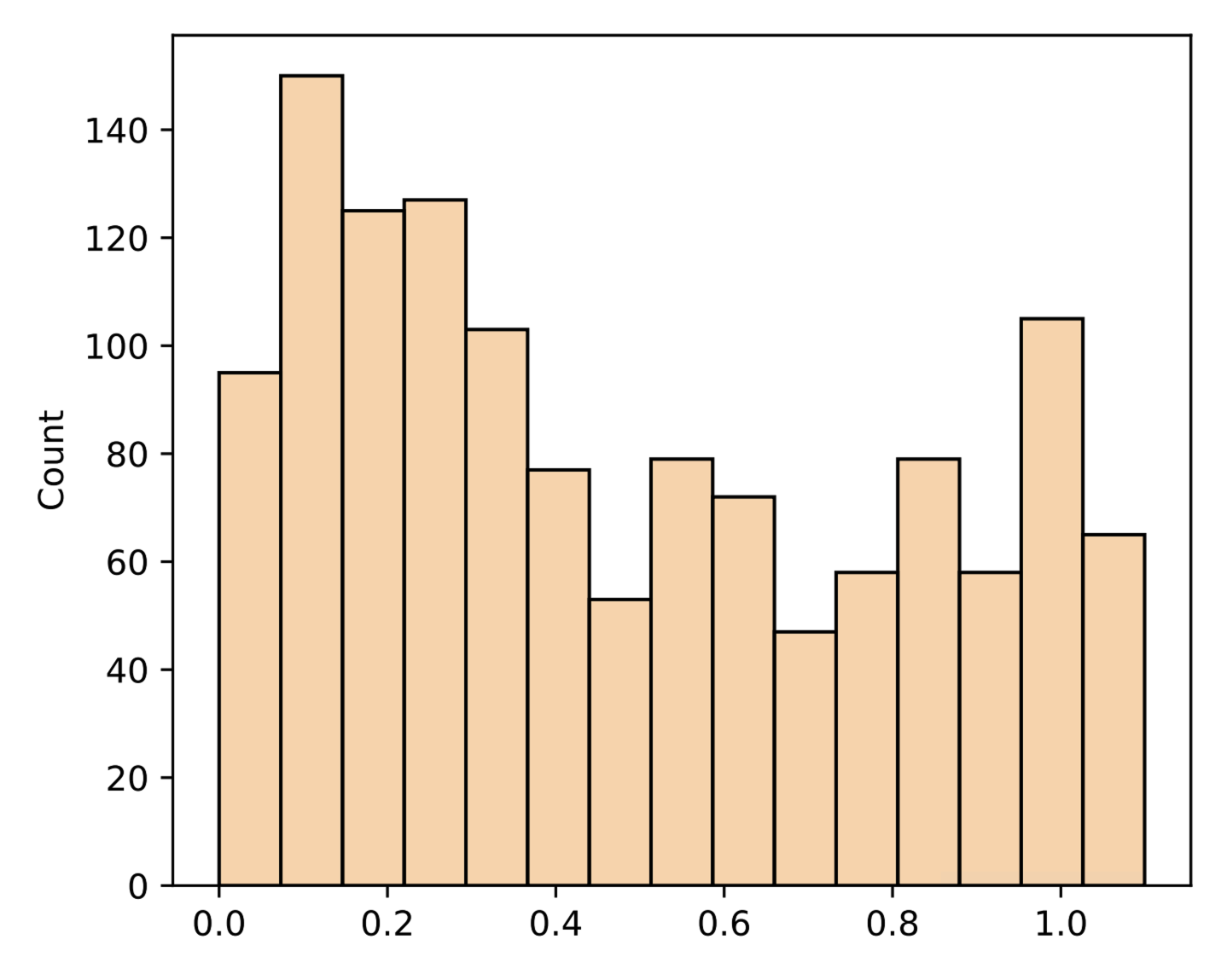} \\
  \end{minipage}
  \vspace{1em}
  \caption[
Comparison of RR Distributions and Feature Space Overlap
]{
Comparison of recovery rate distributions and feature space overlap.
}
  \label{fig:rr_summary_and_venn}
\end{figure}
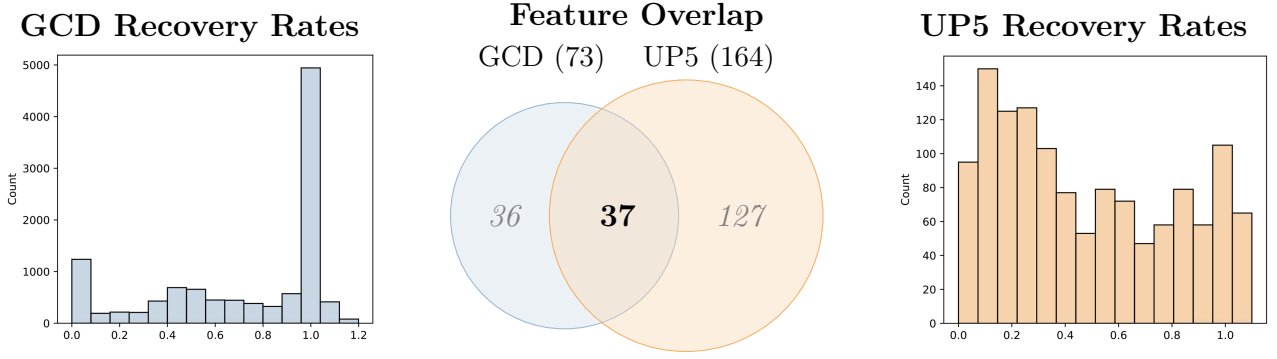

Figure~\ref{fig:rr_summary_and_venn} compares the \emph{post-preprocessing} RR distributions and visualizes the degree of feature-space overlap. GCD exhibits a pronounced bimodal structure with peaks near 0 and 1, whereas UP5 places more mass on low recoveries (mode $\approx 0.1$). Table~\ref{tab:dist_gaps} summarizes these gaps in terms of mean, dispersion, and skewness.

Figure~\ref{fig:rr_summary_and_venn} shows that only 37 predictors overlap between the preprocessed GCD (73 features) and UP5 (164 features) datasets (before dummy coding). This overlap is sufficient to define shared-feature baselines and provides an anchor set of covariates for cross-domain alignment. However, it does not eliminate the key challenge in our setting: most predictors are domain-specific, so restricting models to the shared subset discards substantial information and, by itself, does not resolve feature mismatch. This motivates methods that can transfer across partially overlapping feature spaces while leveraging both shared and portfolio-specific covariates.

\vspace{-1.5em}

\begin{table}[ht!]
  \centering
  \caption[
Key Distributional Gaps Between Source and Target Domains
]{
Key distributional gaps between source and target domains.
}
  \label{tab:dist_gaps}
  \begin{tabular}{lccc}
    \toprule
    & \textbf{GCD} & \textbf{UP5} & $\Delta$ (Target$-$Source)\\
    \midrule
    RR mean & 0.72 & 0.48 & -0.24\\
    RR standard deviation & 0.36 & 0.33 & -0.03\\
    RR skewness & -0.85 & 0.33 & +1.18\\
    \bottomrule
  \end{tabular}
\end{table}
\vspace{-1em}

\begin{figure}[ht!]
  \centering
  \begin{subfigure}{0.4\textwidth}
    \includegraphics[width=\textwidth]{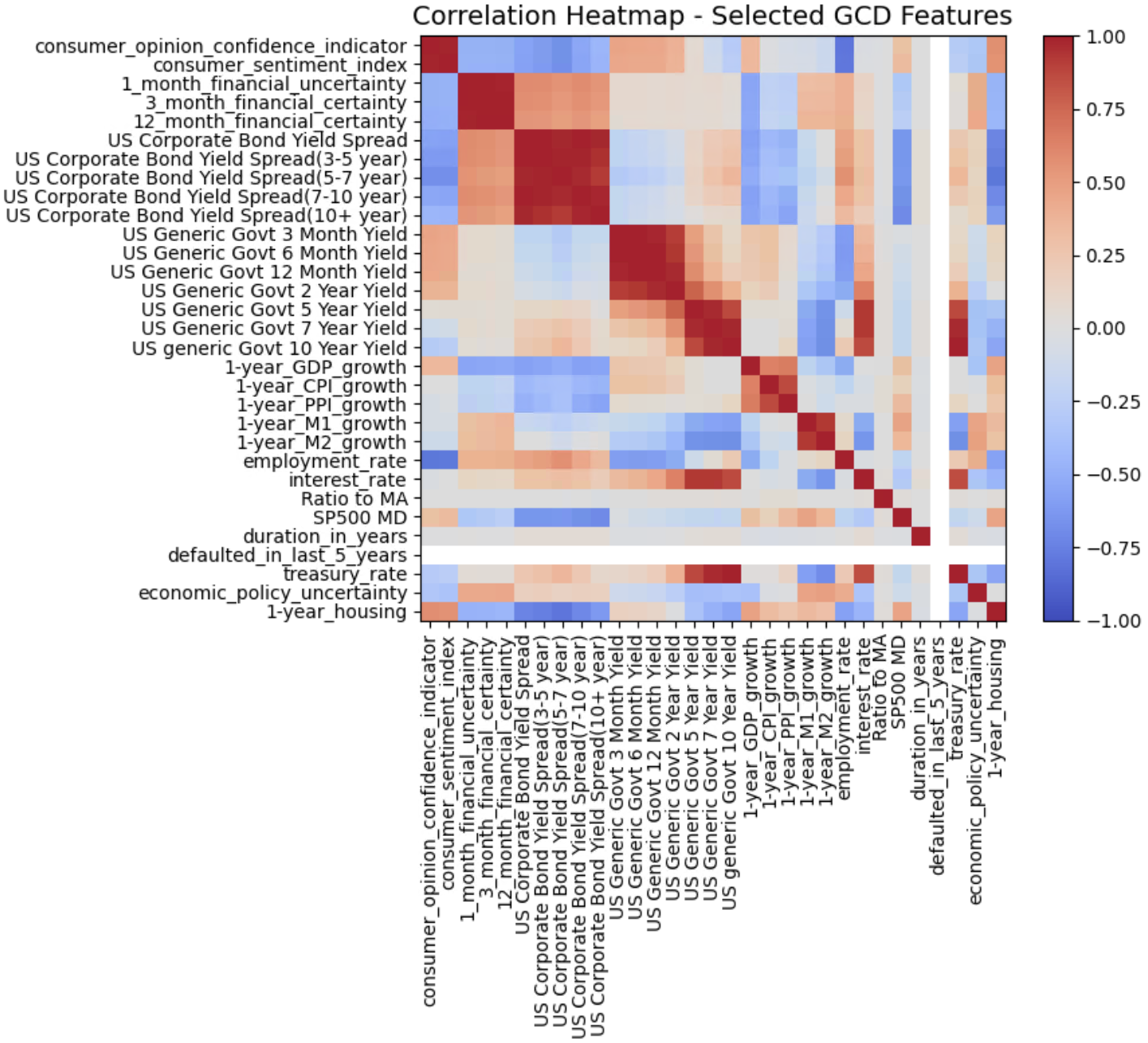}
    \caption{GCD}
    \label{subfig:corr_heatmap_gcd}
  \end{subfigure}
  \hfill
  \begin{subfigure}{0.4\textwidth}
    \includegraphics[width=\textwidth]{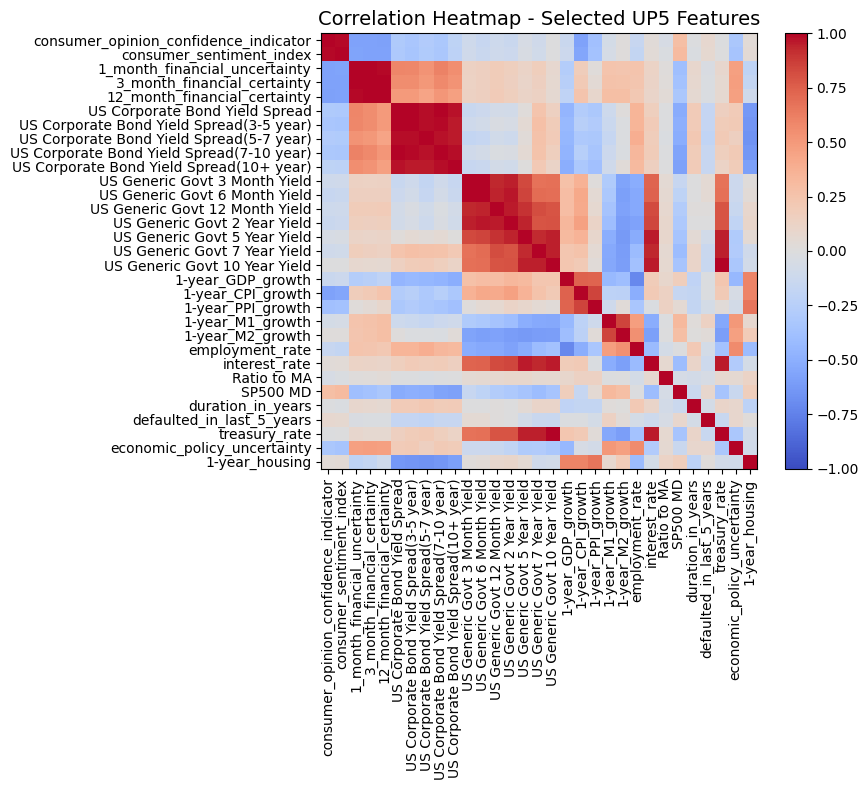}
    \caption{UP5}
    \label{subfig:corr_heatmap_up5}
  \end{subfigure}
  \caption[
Correlation Heatmaps of Macro-Financial Features
]{
Correlation heatmaps (31$\times$31) of common macro-financial features
in GCD and UP5.
}
  \label{fig:corr_heatmap}
\end{figure}

To better understand cross-domain dependencies among continuous covariates, Figure~\ref{fig:corr_heatmap} presents correlation heatmaps for the 31 shared, numerical macro-financial variables in both domains, all measured on comparable scales. We do not compute a single correlation matrix over all shared predictors because several of them %overlapping variables 
are categorical %or encoded as indicator groups, for which Pearson correlations are not meaningful without adopting alternative association measures \citep[e.g.,][]{Akoglu2018}. 
Within the subset of shared macro-financial variables, both domains exhibit qualitatively similar correlation structures, particularly among interest rates, yield spreads, and macro sentiment indicators, suggesting that broad macro-level co-movements are relatively stable across domains. This pattern does not imply that feature relationships are identical in general. Portfolio composition and domain-specific predictors may induce additional differences not captured by the macro-financial subset. This is shown by, for example, notable differences in the marginal RR distributions between GCD and UP5 (see Figure~\ref{fig:rr_summary_and_venn} and Table~\ref{tab:dist_gaps}).

\vspace{-1em}

To clarify feature dimensionality, we distinguish between the conceptual feature space (pre-dummy) and the numerical design matrices used by baselines that require fixed-dimensional inputs. In UP5, the 9 categorical variables are expanded into dummy indicators for such models, yielding a higher-dimensional input representation. For GCD, we avoid a full one-hot expansion because many categorical fields are domain-specific and would introduce a large set of non-overlapping indicators without improving cross-domain comparability. Thus, we dummy-encode only the shared categorical predictors that also occur in UP5 and map them to aligned indicator groups across domains. This design keeps shared categorical information comparable while preventing an unnecessary increase in dimensionality from GCD-only categories; models that natively handle categorical embeddings (including our FT--Transformer) operate on the pre-dummy representation and are, therefore, not dependent on the partial expansion.

\subsection{Real-Data Experimental Design}
\label{subsec:real_data_experiments}

We conduct four complementary real-data configurations that progressively exercise the modeling and transfer mechanisms while keeping preprocessing minimal and comparable across models.  
Each configuration corresponds to one experimental step, aligning with the sequence of results in Section~\ref{sec:results-rr}.

\begin{enumerate}[label=\textbf{Step~\arabic*:}, leftmargin=4em, itemsep=0pt, topsep=0pt]

\item \textbf{Target-Baseline density modeling on UP5}  
FT--MDN--T is trained solely on UP5 using a standard train--validation split and \(K{=}2\) mixture components.  
This baseline examines how the model captures loan-level RR densities and reproduces the overall portfolio distribution without involving any transfer or alternative encoding methods.  
The step establishes the reference performance against which all subsequent transfer- and representation-based improvements are measured.

\item \textbf{Categorical encoding ablation on UP5}  
Two otherwise identical FT--MDN--T variants are compared: one using native categorical embeddings (embedding dimension \(d_\text{cat}{=}d\)) and one using one-hot/dummy expansion.  
Backbone, head, splits, and optimization settings are matched to isolate the impact of categorical representation.  
We report the \(R^2\) to quantify accuracy differences arising purely from encoding choice.  
This step isolates the effect of representation choices so that later transfer experiments do not conflate schema alignment with encoding quality.

\item \textbf{Within-UP5 heterogeneous schema expansion}  
To evaluate warm-starting when new features become available, we first pretrain for up to 100~epochs on the 37-feature subset (columns overlapping with GCD) with early stopping, then expand to the full 317-feature schema and fine-tune for another 100~epochs.  
During the first \(e_{\mathrm{warm}}{=}10\) epochs of fine-tuning, shared-token embeddings and the first Transformer block remain frozen for stability; afterwards all parameters are unfrozen and optimized jointly with a \(10\times\) smaller learning rate.  
A model trained from scratch on the full schema with the same training budget serves as a baseline.  
This step provides an in-domain test of the heterogeneous-schema mechanism before applying it across portfolios.

\item \textbf{Cross-portfolio transfer from GCD to UP5}  
We test multiple transfer regimes, including the following:  
(\emph{i})~\emph{shared$\rightarrow$full}, where pretraining uses GCD restricted to the 37 features shared with UP5 and fine-tuning expands to the full 317-feature UP5 schema; and  
(\emph{ii})~\emph{full$\rightarrow$full}, where pretraining uses the entire GCD schema and source-only features are masked during transfer.  
Target training size is varied with \(n_t\in\{100,300,500,1000,\text{all}\}\).  
For each \(n_t\), results are averaged over random subsets and seeds (15 different runs in total), and reported as mean~\(\pm\)~standard error.  
This final step evaluates the full heterogeneous-schema transfer pipeline under real cross-portfolio distribution shifts.
\end{enumerate}

The real-data experiments progress from in-domain density modeling (Step 1) to encoding robustness (Step 2), schema transfer within a domain (Step 3), and cross-portfolio transfer across domains (Step 4).  
The subsequent simulation study (Step 5) extends these experiments under controlled shift and heterogeneity conditions.

\subsection[Step 1: Added value of distributional modeling]{Step 1: Added value of distributional modeling (UP5, Target-only)}
\label{subsec:dist_value}

In the first experiment, we illustrate how FT--MDN--T models the full conditional distribution of recovery rates rather than a single point estimate.
We train the model on the UP5 dataset using a standard train--test split. No transfer or baseline models are used at this stage---the goal is purely to inspect how the MDN head represents RR uncertainty.

\begin{figure}[ht!]
    \centering
    \includegraphics[width=0.45\linewidth]{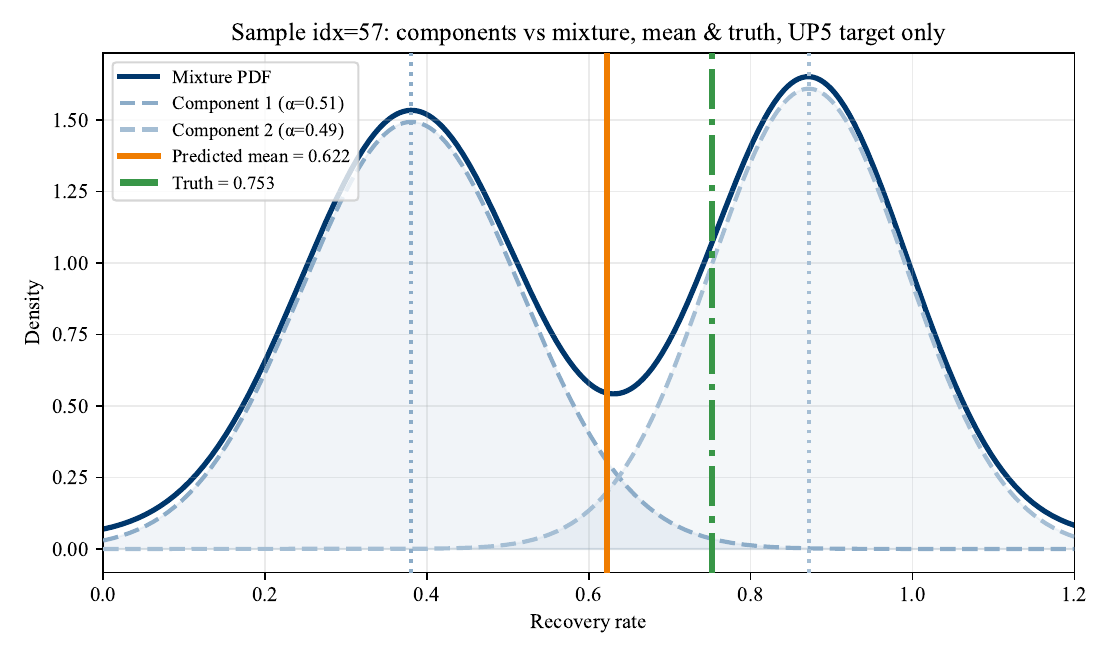}
    \includegraphics[width=0.45\linewidth]{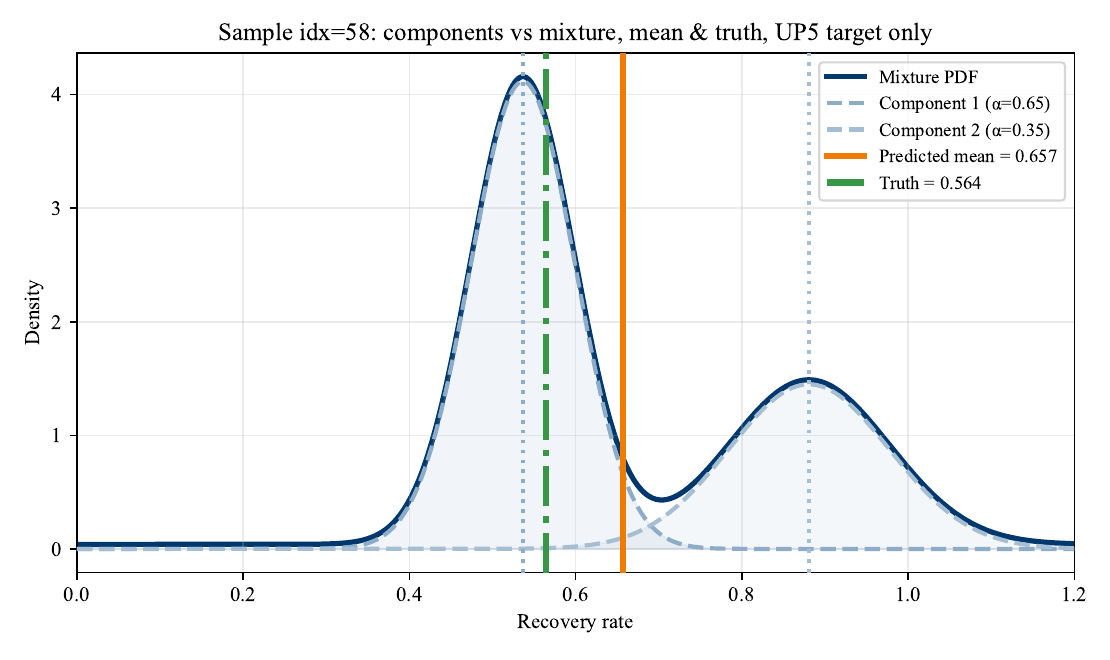}\\
    \includegraphics[width=0.45\linewidth]{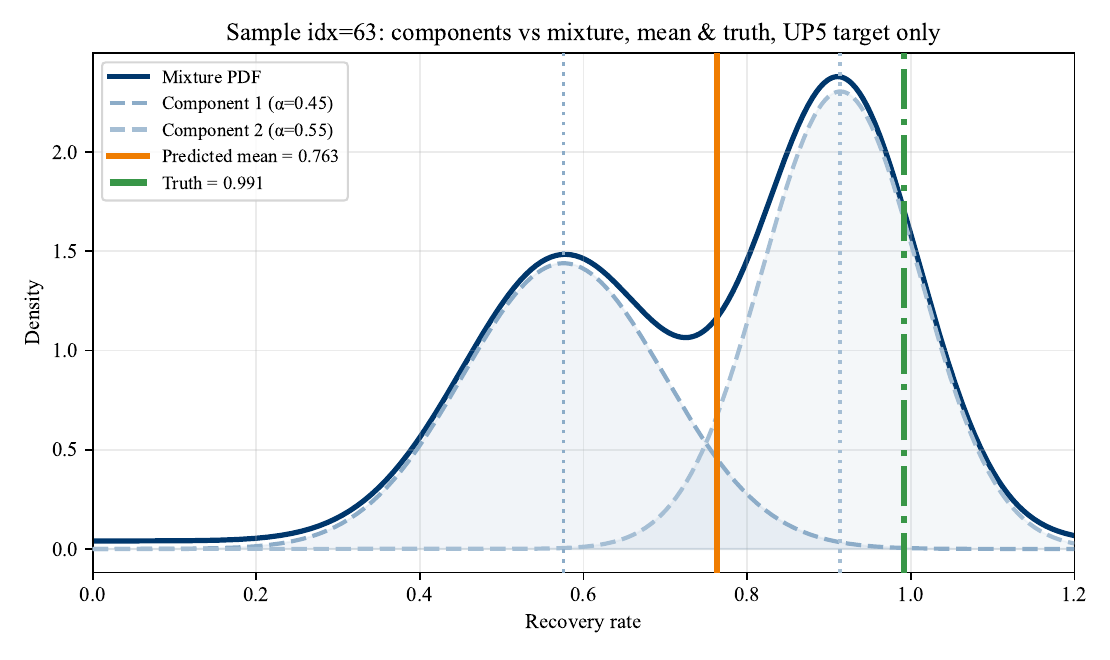}
    \includegraphics[width=0.45\linewidth]{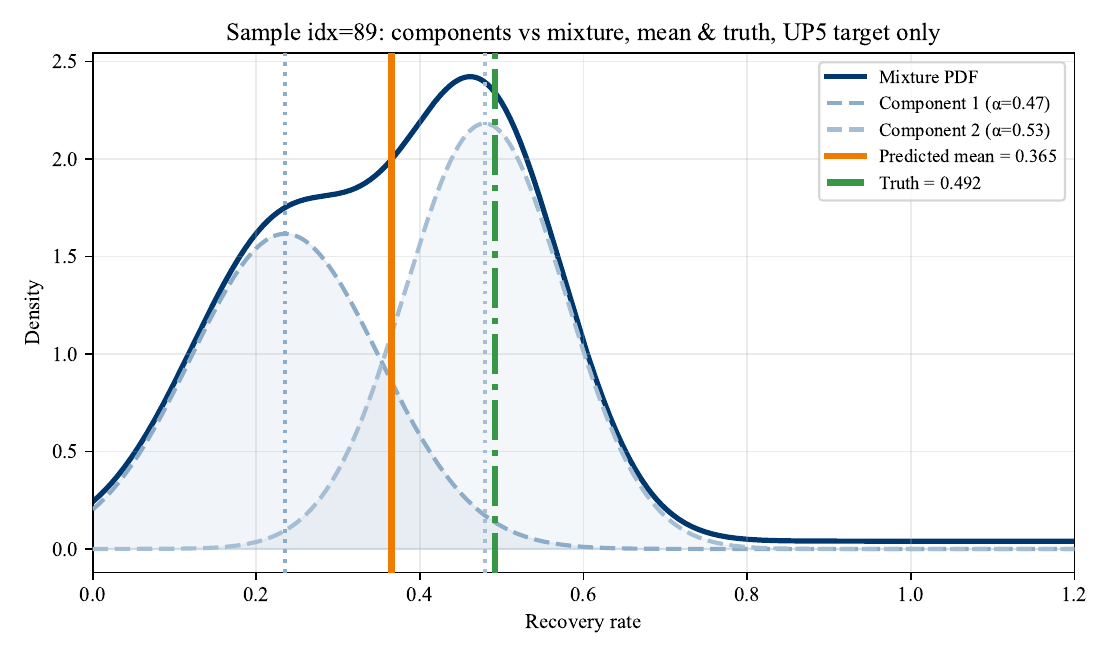}
    \caption[
Single-Loan Recovery Distributions (UP5)
]{
Single-loan recovery distributions predicted by FT--MDN--T on UP5.
Mixture components capture bimodality that is lost when only the mean is considered.
}
    \label{fig:mdn_samples}
\end{figure}

Figure~\ref{fig:mdn_samples} shows several individual test loans, each with its predicted mixture distribution. The two Gaussian components correspond to the learned modes of the RR distribution, and their weighted sum yields the overall predictive density for each loan. These densities often exhibit a clear bimodal shape, with one mode near low recoveries and another near high recoveries. The corresponding point predictions (mixture means) frequently collapse toward central values, masking the underlying multimodal structure. The distributional outputs, by contrast, expose the distinct recovery regimes that such single-valued predictions cannot represent.

\begin{figure}[ht!]
    \centering
    \includegraphics[width=0.8\linewidth, trim=0 0 0 40, clip]{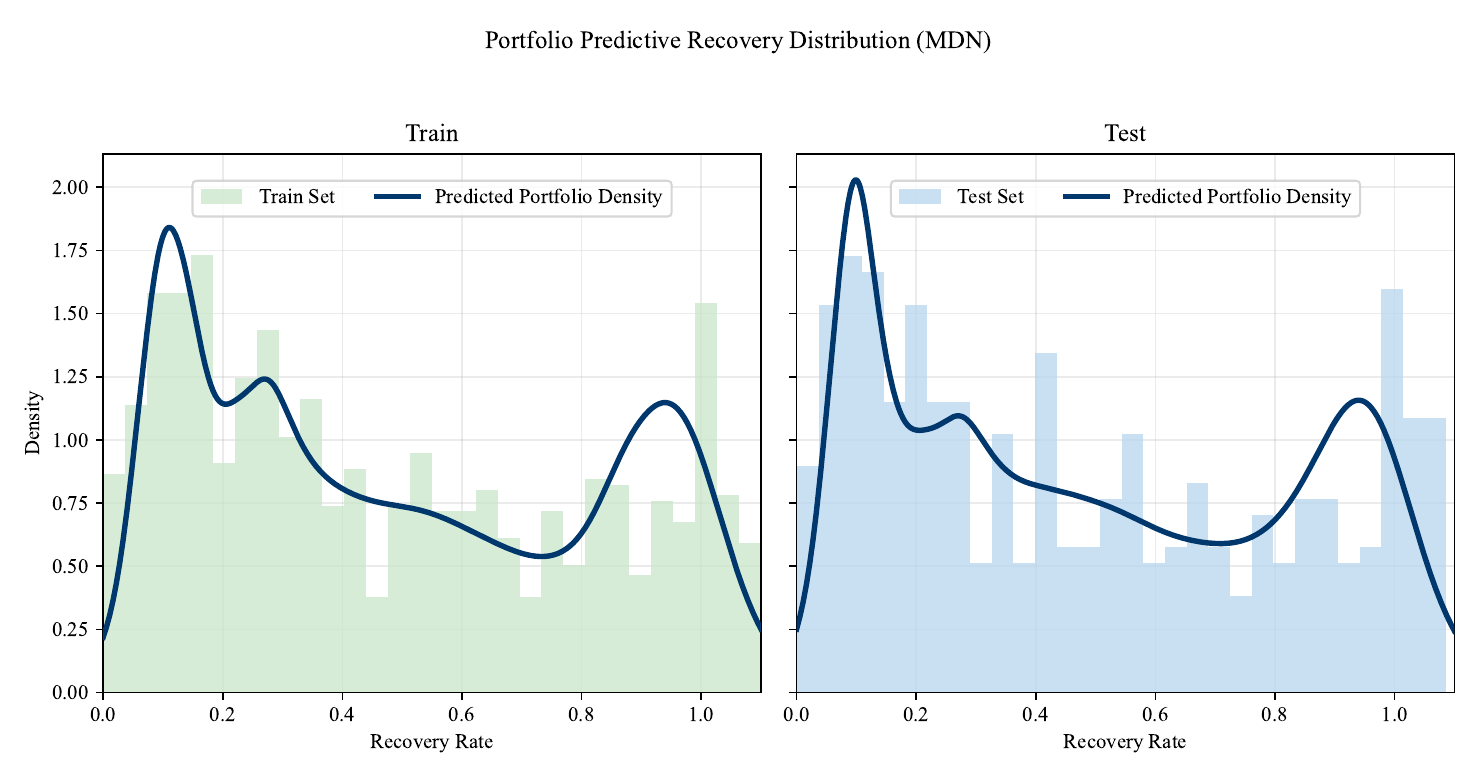}
    \caption[
UP5 Portfolio-Level Recovery Distributions
]{
UP5 portfolio-level recovery distributions and FT--MDN--T based density estimate.
%FT--MDN--T captures bimodality, unlike point-estimation models.
}
    \label{fig:up5_density}
\end{figure}

To move from individual to aggregate behavior, we average all predicted densities across the test portfolio.  
Figure~\ref{fig:up5_density} compares this aggregated (mixture-of-mixtures) prediction with the empirical recovery distribution of UP5.  
The close visual alignment indicates that FT--MDN--T reproduces the overall shape of the characteristic bimodal distribution.  
This ability to recover realistic RR distributions is particularly valuable in risk applications such as provisioning or stress testing, where decision makers care about the entire loss distribution---including tail risks and multi-regime behavior---rather than a single summary statistic such as $R^2$ or MAE.  
By capturing this richer structure, distributional modeling provides a more informative and operationally relevant view of portfolio risk that point estimation models.

\subsection{Step 2: Encoding Categorical Features}
\label{subsec:categoricals}
In RR modeling, tabular datasets typically contain a mix of numerical and categorical variables, such as industry sector, region, rating class, or collateral type. Handling these categorical features effectively is essential for predictive performance and model robustness, especially when transferring across portfolios with differing category sets.

This aspect becomes particularly critical in TL settings, where categorical feature spaces rarely align perfectly across portfolios. In practice, source and target datasets often differ not only in the number of categories but also in their semantic composition, for example, industry taxonomies may be more granular in one portfolio, regional codes may change, or rating classes may be merged or split over time. Under such conditions, conventional dummy encoding yields a brittle feature representation: categories absent in the target domain must be dropped, newly emerging categories cannot be reused from pretraining, and even partially overlapping category sets can lead to structural breaks in the input space.

By contrast, embedding-based representations offer a more flexible mechanism for cross-portfolio transfer. Learned embeddings decouple categorical identity from fixed column positions, allowing pretrained representations to be reused, extended, or partially masked when category sets differ between source and target. This makes categorical embeddings a natural complement to schema-flexible transfer learning, as they enable the model backbone to preserve and adapt semantic information even when exact category alignment is not possible.

A practical advantage of FT--MDN--T is its seamless treatment of categorical variables through learned embeddings, which transform each category into a dense, trainable representation. This approach eliminates the need for manual one-hot or dummy encoding, which is still commonly applied in traditional tabular models such as XGBoost or Random Forest. Such dummy expansion can dramatically increase feature dimensionality, introduce sparsity, and complicate feature alignment when schemas differ across datasets.

\begin{figure}[ht!]
    \centering
    \includegraphics[width=0.8\linewidth, trim=0 0 0 20, clip]{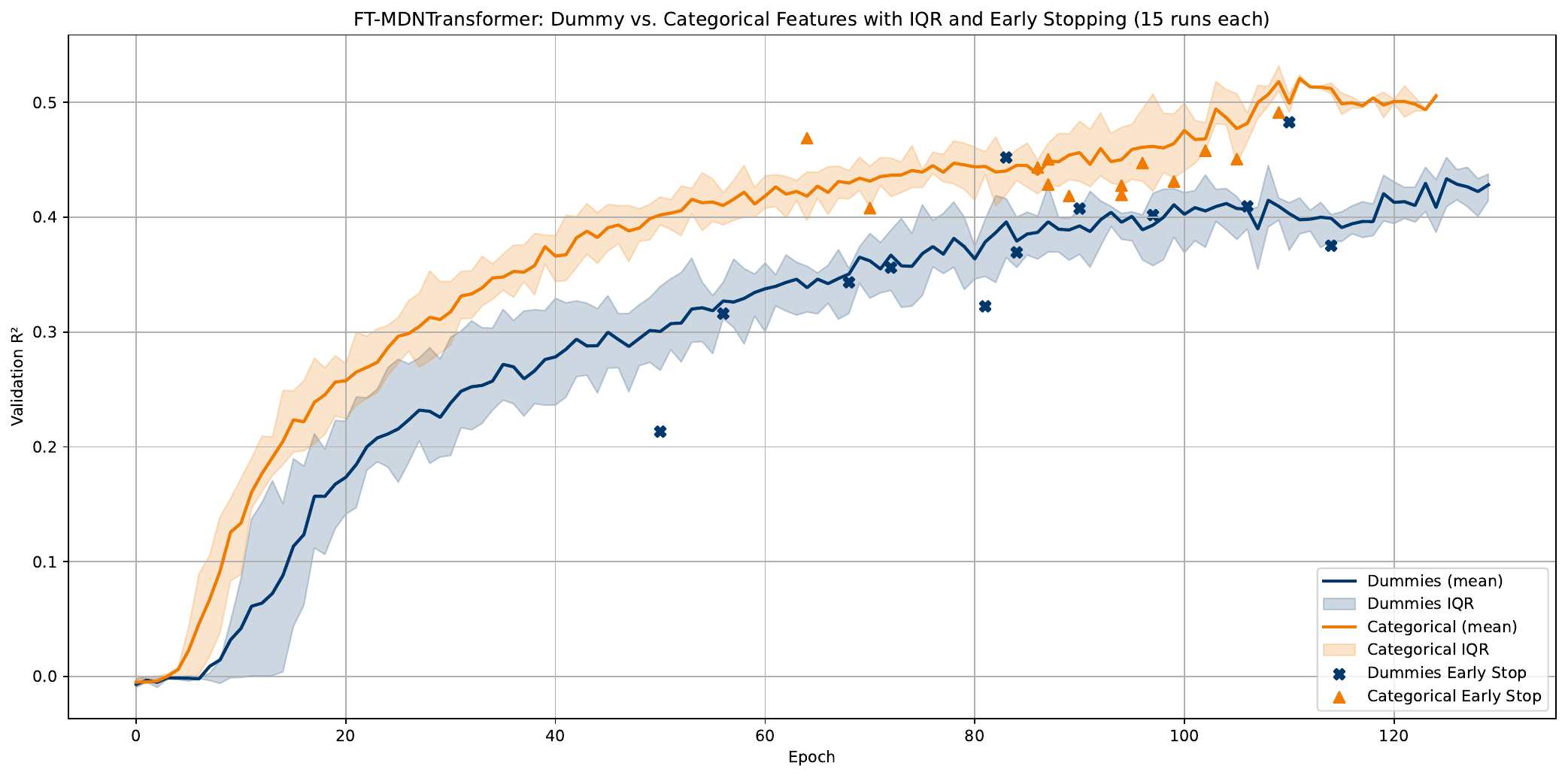}
    \caption[
Categorical Embeddings vs. Dummy Encoding (UP5)
]{
Categorical embeddings versus dummy encoding on UP5.
FT--MDN--T benefits from native embeddings, unlike tree-based baselines.
}
    \label{fig:cat_vs_dummy}
\end{figure}

To quantify the effect of embedding-based encoding, we train two FT--MDN--T variants on the UP5 dataset using identical settings: one with native categorical embeddings and one with manually expanded dummy variables for the same categorical fields. Figure~\ref{fig:cat_vs_dummy} shows the validation \(R^2\) of both model variants over training epochs. Solid lines report the average performance across 15 runs with different random seeds, while shaded areas represent the interquartile range (IQR). 
Markers indicate the epochs at which early stopping is triggered, based on the absence of further improvements in validation performance.
The embedding-based variant consistently achieves higher performance throughout training and reaches stable performance earlier, as reflected by earlier and more tightly clustered early-stopping points. 
In contrast, the dummy-encoded model improves more slowly and exhibits greater variability across runs, resulting in later, more dispersed stopping epochs. 
Overall, these results indicate that native categorical embeddings not only improve predictive accuracy but also lead to more stable and reliable training behavior.

This result reinforces a broader methodological advantage: FT--MDN--T removes one of the key bottlenecks in credit risk modeling---the manual handling of categorical features---while preserving accuracy and adaptability across heterogeneous datasets.

\subsection{Step 3: Proof-of-Concept for Heterogeneous Feature Transfer}
\label{subsec:hetero_up5}

Before studying cross-portfolio transfer under simultaneous distributional shift and schema mismatch, we first isolate the role of feature heterogeneity in a controlled setting. 
Specifically, we examine whether FT--MDN--T can reuse and extend representations learned on a limited feature subset when additional target-specific features become available, \emph{without} introducing any distributional shift. 
This experiment, therefore, serves as a TL proof-of-concept that focuses exclusively on schema expansion, abstracting from the confounding effects of unavoidable feature or label drift in real cross-portfolio settings.

Concretely, we conduct all experiments within the UP5 domain. 
We first pretrain FT--MDN--T for 100 epochs using only a subset of 37 features, corresponding to those shared between GCD and UP5. 
After this pretraining phase, we expand the input space to the full UP5 feature schema with 317 features and continue training (fine-tuning) for another 100 epochs under identical optimization settings. 
Because both phases use the same underlying data distribution, this setup allows us to attribute any performance gains solely to the model’s ability to incorporate new features, rather than to changes in the target domain itself.

\begin{figure}[ht!]
    \centering
    \includegraphics[width=0.85\linewidth, trim=0 0 0 25, clip]{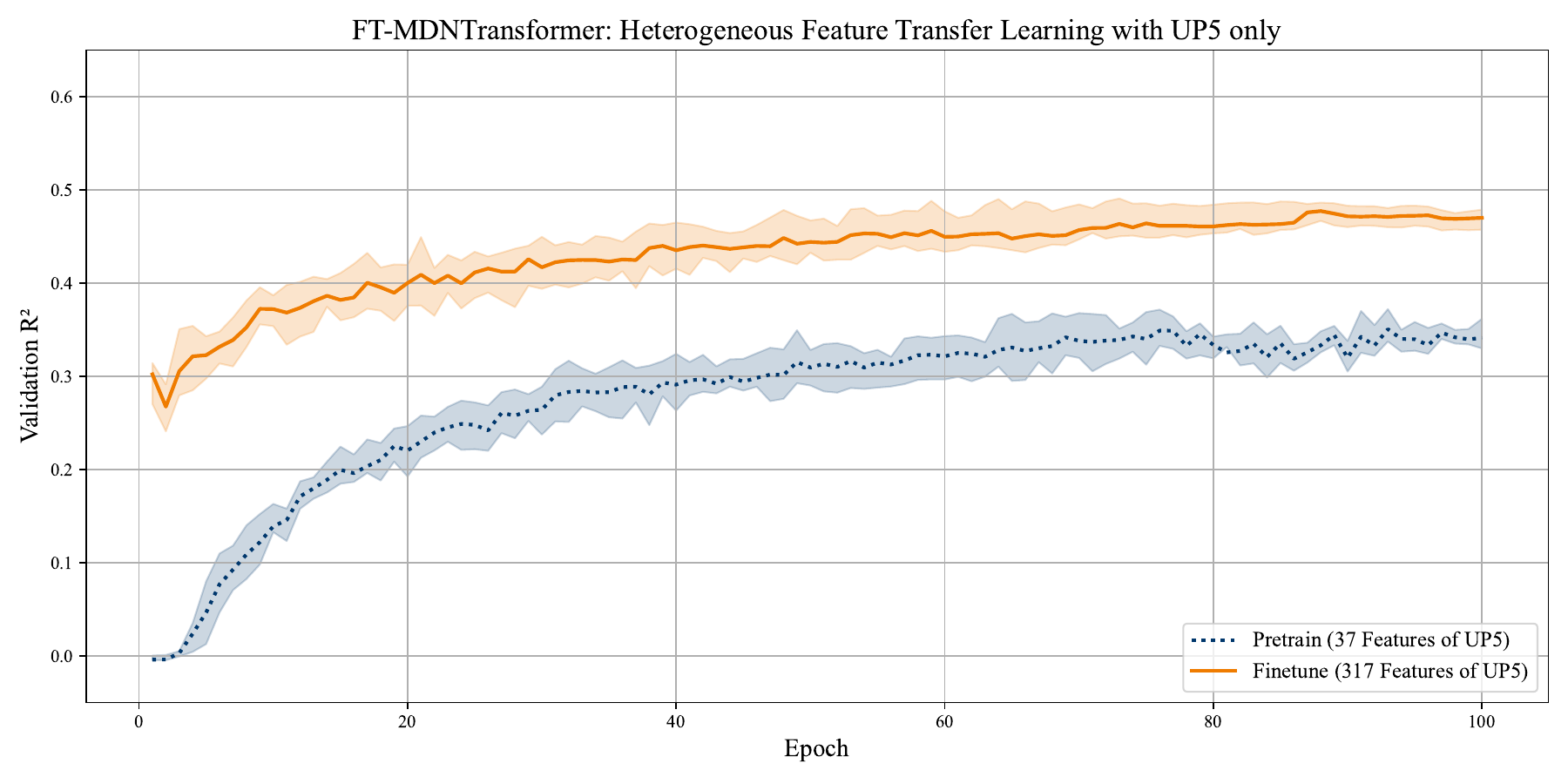}
    \caption[
Validation $R^2$ on UP5
]{
Validation $R^2$ on UP5 when expanding from 37 to 317 features.
Pretraining accelerates fine-tuning and surpasses the scratch baseline.
}
    \label{fig:up5_shared_to_full}
\end{figure}
\vspace{-1em}

Figure~\ref{fig:up5_shared_to_full} reports validation \(R^2\) over training epochs. 
The blue dotted line shows performance during pretraining on the limited feature subset, which starts near zero and gradually rises to about \(R^2 \approx 0.3\), where it stabilizes and shows no further systematic improvement. 
When fine-tuning begins (orange line), the model immediately continues from this level rather than restarting at zero, indicating that representations learned on the shared feature subset remain useful after the schema is expanded. 
Here, epoch~1 on the orange curve corresponds to the first fine-tuning epoch after the initial 100 epochs of pretraining. 
As training proceeds, performance further improves to around \(R^2 \approx 0.47\) and again converges to a stable plateau, clearly exceeding the level achieved by training on the reduced feature set alone.

This result provides a first empirical confirmation that FT--MDN--T can transfer knowledge across heterogeneous feature spaces in a setting without distributional shifts.
Rather than relearning from scratch, the model effectively reuses representations learned on a shared feature subset and integrates additional target-only features. 
By explicitly eliminating distributional drift, this setup isolates feature-space heterogeneity as the sole challenge and shows that FT--MDN--T can accommodate incremental schema expansion. 
This controlled result lays the groundwork for the subsequent GCD~$\rightarrow$~UP5 analysis, where feature mismatch and distributional shift occur jointly and cannot be disentangled in real data.

\subsection[Step 4: Cross-Portfolio Transfer]{Step 4: Cross-Portfolio Transfer (GCD → UP5)}
\label{subsec:gcd_up5}

\begin{figure}[ht!]
\centering
\includegraphics[width=0.45\linewidth]{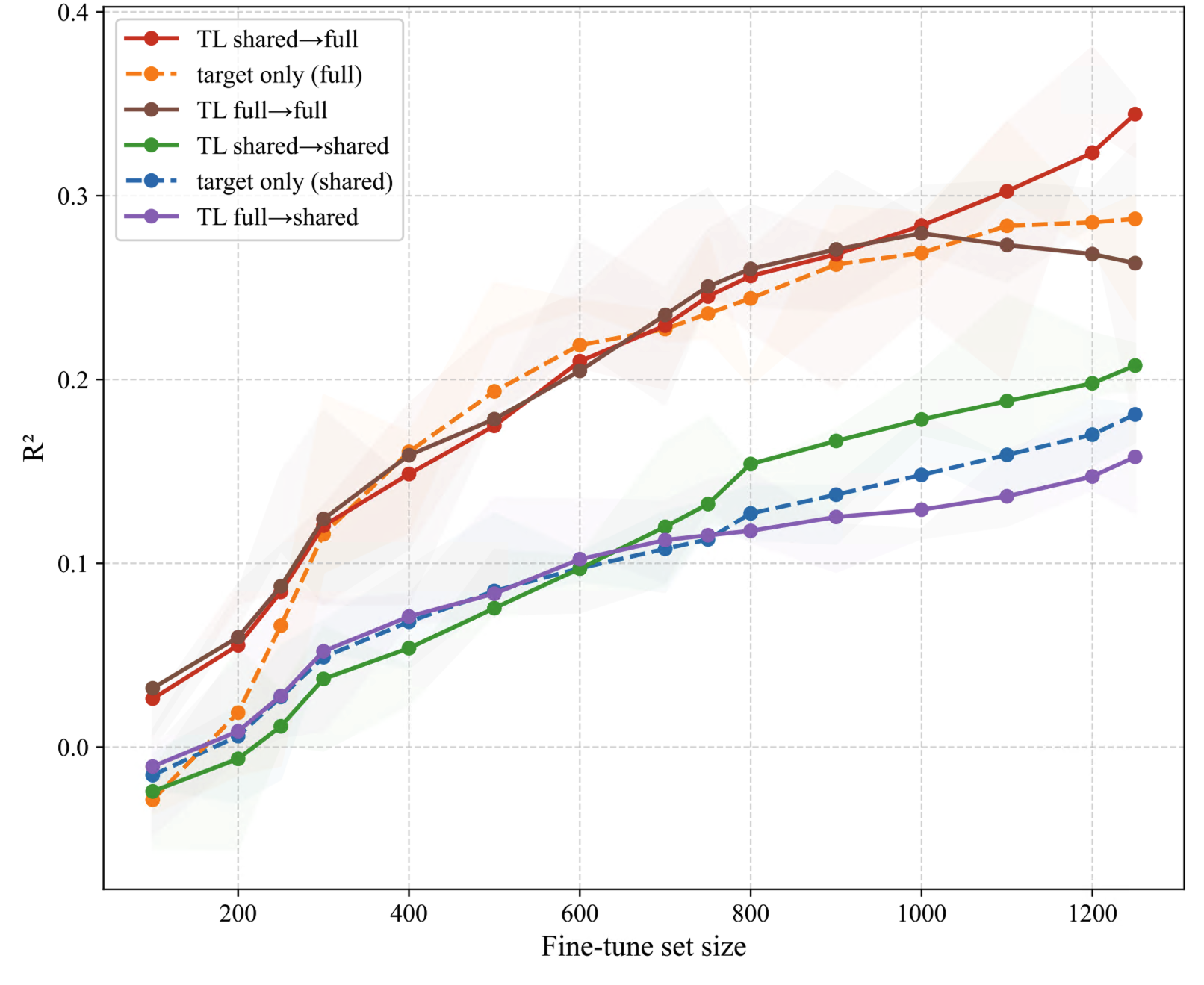}
\includegraphics[width=0.45\linewidth]{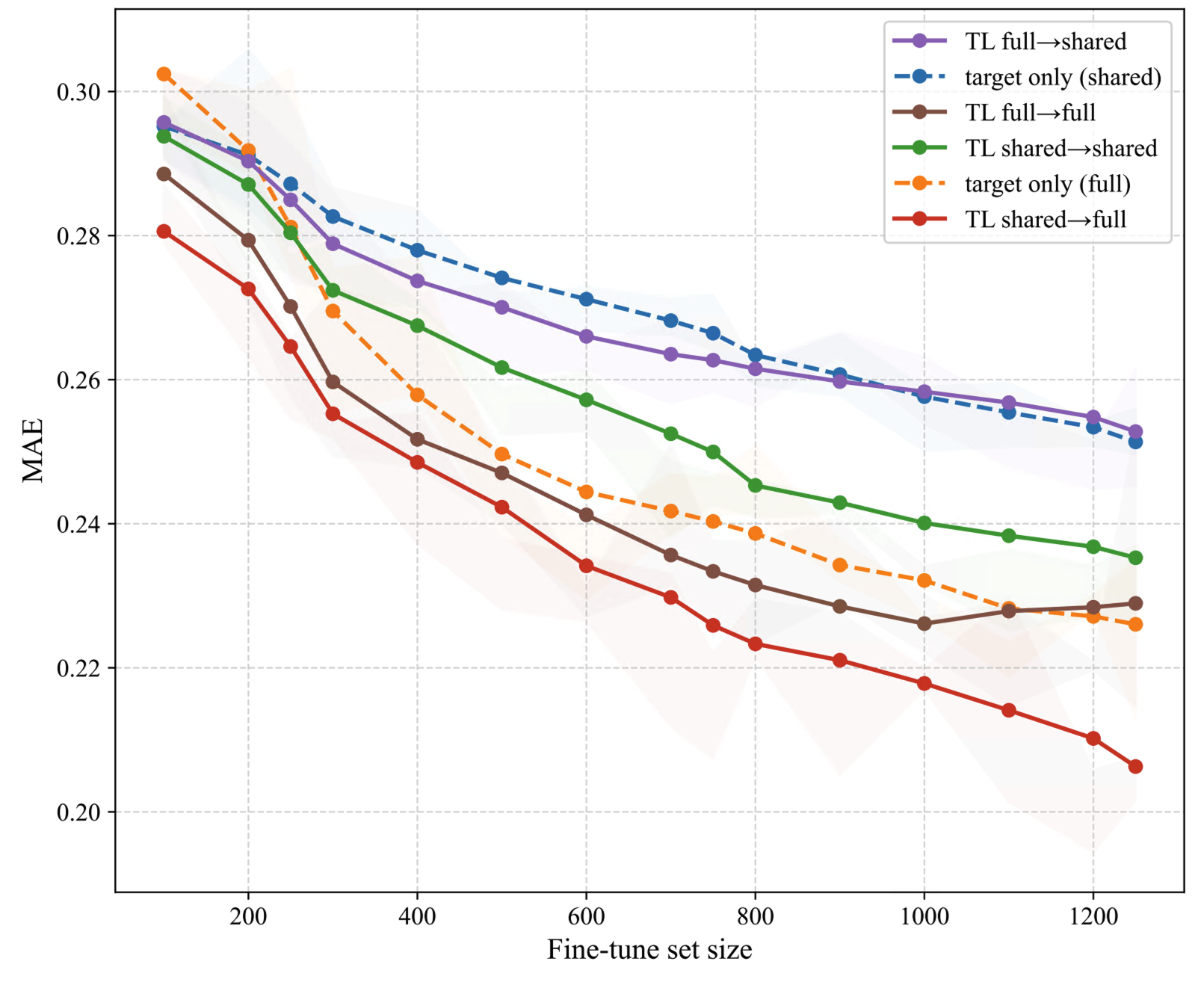}
\caption[
Cross-Portfolio Transfer: GCD to UP5
]{
Cross-portfolio transfer from GCD to UP5.
Left: $R^2$. Right: MAE.
TL shared$\rightarrow$full shows strong results, followed by
TL full$\rightarrow$full.
}
\label{fig:gcd_up5_transfer}
\end{figure}

We next examine the most challenging setting in our study: transferring knowledge from a loans (GCD) to a bonds dataset (UP5). 
This scenario mirrors the practical challenges of cross-portfolio transfers, as the two domains differ markedly in contractual structures, collateralization, geography, and feature availability.
The experiment uses all observations from GCD to pretrain FT--MDN--T under two feature regimes: one using the full GCD feature set and one restricted to the 37 features shared with UP5. 
These pretrained models are then fine-tuned on UP5, either using the shared 37 features or the full 317-feature schema, while varying the number of UP5 training samples. Thus, the x-axis in Figure~\ref{fig:gcd_up5_transfer} represents increasing target data availability. All curves report averaged validation metrics across multiple random subsets and seeds (with shaded areas indicating variability across runs). 
Here, ``full'' always refers to the feature set, not to the number of observations.

\vspace{-1em}

The results show several consistent patterns. 
When the target dataset is very small, transfer learning substantially improves both \(R^2\) and MAE relative to target-only training. 
The best performance is achieved by \emph{TL shared$\rightarrow$full}, where the model is pretrained on shared GCD features and fine-tuned with the full UP5 feature schema, demonstrating that a model pretrained on the shared subset can incorporate additional UP5-only features during fine-tuning.

\emph{TL full$\rightarrow$full} also performs competitively but does not surpass \emph{TL shared$\rightarrow$full}. 
One might expect that pretraining on a richer feature set would provide more transferable knowledge. However, in this setting, features present in GCD but absent in UP5 appear to bias the learned model during pretraining, potentially hindering adaptation when these features disappear at fine-tuning. As a result, the model benefits more from pretraining on the cleanly aligned, shared subset before expanding to the full UP5 schema. This finding suggests that successful transfer across heterogeneous domains depends not only on the number of features used during pretraining, but also on how well the source feature structure aligns with the target schema. 
Pretraining on features that later disappear in the target domain can hinder adaptation, whereas pretraining on a cleanly aligned shared subset yields a more transferable initialization. We will revisit this phenomenon in the simulation study, where feature-space relations (source $\subset$ target, target $\subset$ source, partial overlap) are varied explicitly to isolate these effects. As UP5 training data increase, the relative advantage of transfer diminishes, and target-only models gradually close the gap. Notably, this occurs already at relatively small target sample sizes, indicating that FT--MDN--T can learn effectively even in target-only settings with limited data. 
Given the complexity of Transformer-based models, this level of sample efficiency is noteworthy and suggests that transfer learning provides its largest benefits primarily in the most data-scarce regimes. Overall, these results demonstrate that FT--MDN--T adapts flexibly across heterogeneous feature spaces: it can incorporate new target-only features and remain functional when source-only features disappear, while highlighting that extreme domain mismatch (loans vs.\ bonds) inherently limits transfer gains.
While the real-data study shows how FT--MDN--T behaves under realistic portfolio differences, it cannot disentangle individual sources of heterogeneity and shift. The following simulation study addresses this limitation by isolating these mechanisms under controlled conditions.

\section{Simulation Study and Robustness}
\label{sec:results-rr}

This section extends the real-data analysis with controlled Monte Carlo simulations to assess robustness under distributional shift and schema heterogeneity. While the real-data study evaluates transfer under naturally occurring portfolio differences, the simulations isolate individual shift mechanisms and feature-overlap regimes that cannot be disentangled empirically.

\subsection{Simulated Loan Recovery Datasets}

To systematically probe the mechanisms underlying the real-data results, we construct synthetic loan recovery datasets that reproduce key empirical properties while allowing explicit control over distributional shifts.
A direct empirical evaluation of cross-domain transfer in loan recovery modeling is challenging not only because publicly available credit-recovery datasets are scarce, but also because real-world datasets offer limited control over their distributional characteristics. In practice, observed data do not permit systematic manipulation of shifts---such as targeted adjustments to covariates, conditional relationships, or label distributions---that would be required to assess model behavior rigorously. To complement our empirical analysis, we adopt a simulation-based approach to generate synthetic yet realistic loan-recovery datasets. These datasets replicate key empirical patterns, such as bimodal RR distributions and plausible feature correlations, while giving us the flexibility to impose distinct types of distributional shifts in a controlled manner.

The proposed simulation-based framework provides several complementary strengths. First, because the real GCD/UP5 datasets only partially overlap in their feature sets, differences in available variables or measurement definitions may influence a purely empirical cross-dataset experiment. Simulated data allow us to abstract away from such practical constraints. Second, in real data, different forms of drift (marginal shift, covariate shift, conditional shift, label-noise change, etc.) typically co-occur, which makes it difficult to disentangle their individual effects on model performance. Third, synthetic datasets allow us to isolate and vary specific shift types---for example, by modifying marginal distributions, altering conditional relationships, or scaling noise---enabling systematic ablation and sensitivity studies beyond what real data alone can reveal.
%SL: your fourth advantage exists, but we do not exploit this potential
%Fourth, simulations enable the controlled inclusion of rare or extreme cases that may be underrepresented in empirical datasets, thereby supporting more comprehensive stress tests of model behavior.
Finally, the simulation framework is fully reproducible and publicly available, enabling consistent and transparent comparisons across methods and experimental settings that would be difficult to achieve with proprietary real-world datasets.

\subsubsection{Generating Recovery Rates}
A key empirical finding from datasets such as GCD is the presence of distinctly \emph{bimodal} RR distributions, with substantial probability mass concentrated near zero (typical for unsecured exposures) and near one for secured loans \citep{Loterman2012}. This structure reflects fundamental economic differences in recovery processes, in which the presence of collateral strongly alters both the level and the variability of realized recoveries. To reproduce this behavior, we model RR using a two-component Gaussian mixture model,
\vspace{-0.6em}
\begin{equation*}
R \;\sim\; \pi \cdot \mathcal{N}\bigl(\mu_{\text{secured}}, \sigma_{\text{secured}}^2\bigr)
\;+\; (1-\pi) \cdot \mathcal{N}\bigl(\mu_{\text{unsecured}}, \sigma_{\text{unsecured}}^2\bigr),
\vspace{-0.6em}
\end{equation*}
where \(\pi\) denotes the probability that an exposure is secured and therefore drawn from a high-recovery component with mean \(\mu_{\text{secured}}\) and variance \(\sigma_{\text{secured}}^2\). The remaining mass \(1-\pi\) corresponds to unsecured exposures, which are characterized by systematically lower average recoveries \(\mu_{\text{unsecured}}\) and potentially different variance \(\sigma_{\text{unsecured}}^2\). By explicitly separating secured and unsecured regimes, this formulation allows us to control both the relative prevalence of high- and low-recovery observations and the degree of heterogeneity within each regime, while remaining closely aligned with empirical recovery patterns observed in practice.

To generate RR from the mixture distribution, we draw samples by numerically inverting the cumulative distribution function
\vspace{-0.6em}
\[
F_R(r) = \pi \cdot \Phi_{\text{secured}}(r) + (1 - \pi) \cdot \Phi_{\text{unsecured}}(r),
\vspace{-0.6em}
\]
where \(\Phi(\cdot)\) denotes the Gaussian cumulative distribution function. This procedure yields smooth, continuous draws from the mixture while avoiding discrete component assignments during sampling. All simulated recovery values are subsequently clipped to the range \([0,1]\). In addition, each observation is assigned a binary indicator (\texttt{secured}) that records the underlying mixture component, enabling later analyses to condition explicitly on security status.

\subsubsection{Feature Generation}
Having specified the recovery-generating mechanism, we next construct feature vectors conditional on \(R\), ensuring realistic dependence structures between covariates and recovery outcomes.
A central element of our simulation approach is to model features \(X = (X_1, \dots, X_p)\) \emph{conditional on} the recovery rate \(R\). Here, indices \(j\), \(k\), \(a\), and \(b\) refer to individual features, where \(j\) denotes a generic feature, \(k\) an interaction feature, and \(a,b\) the base features of an interaction. This enables the generation of realistic correlations in which feature values systematically differ between low- and high-recovery observations, mirroring patterns observed in real-world loan datasets \citep{Loterman2012, Yao2017}.
A variety of functional forms for numeric features \(X_j\) are supported, as summarized in Table~\ref{tab:feature-types}.

\begin{table}[ht]
\centering
\small
\caption[
Feature Relationships for Synthetic Data Generation
]{
Common feature relationships for synthetic data generation.
}
\label{tab:feature-types}
\begin{tabularx}{\textwidth}{l l X}
\toprule
\textbf{Type} & \textbf{Functional Form} & \textbf{Example / Comments} \\
\midrule

\textbf{Linear} &
\(\displaystyle X_j = \beta_j R + \text{intercept}_j + \epsilon_j\) &
Captures the proportional effect of recovery. Additive noise \(\epsilon_j\) and skew can be configured. \\[6pt]

\textbf{Nonlinear} &
\(\displaystyle X_j = \beta_j f(R) + \epsilon_j\) &
Supports polynomial, sine, exponential, and sigmoid functions. Reflects non-monotonic relationships. \\[6pt]

\textbf{Custom} &
User-defined function \(X_j = g(R) + \epsilon_j\) &
Arbitrary Python expressions, e.g., \verb|lambda rr: np.square(rr)|. \\[6pt]

\textbf{Interaction} &
\(\displaystyle X_k = f(X_a, X_b)\) &
Feature constructed from other features via multiplication or addition. Reflects derived ratios and indices. \\
\bottomrule
\end{tabularx}
\end{table}
\vspace{-0.6em}

We implement eight synthetic features that collectively represent all four types of Table~\ref{tab:feature-types}, each with varying levels of complexity. Key design principles include:

\begin{enumerate}[itemsep=0pt, topsep=0pt]
    \item \textbf{Base Relationship to \(R\):}  
    Each feature \(X_j\) is constructed as a deterministic transformation of the recovery rate 
    (e.g., linear, polynomial, or sinusoidal), possibly with configurable skew and scale.
    
    \item \textbf{Independent Random Variation:}  
    To avoid perfect correlation with \(R\), we inject a separate noise term 
    \(\epsilon_j \sim \mathcal{N}(0, \sigma_j^2)\) into each feature. 
    This mimics measurement error and latent influences.
    
    \item \textbf{Feature Interactions:}  
    Some variables in real datasets are constructed as functions of others 
    (e.g., the product of two financial indicators). 
    We replicate this structure by allowing interaction features 
    \(X_k = f(X_a, X_b)\) derived from already generated ones.
    
    \item \textbf{Distributional Anchoring:}  
    To increase realism, we loosely calibrate each synthetic feature based on archetypes 
    from the GCD dataset. Features are grouped (e.g., borrower-level, collateral-specific, macroeconomic), 
    and we match empirical characteristics such as variance, skewness, and scaling.
\end{enumerate}

In addition to continuous variables, our framework supports \textbf{categorical features} 
via a softmax-based sampling mechanism. 
For a categorical variable with classes \(c = 1, \dots, C\), class logits are defined as:
\vspace{-1em}
\begin{equation*}
\text{logit}_c = w_{0c} + w_{1c} R + w_{2c} R^2 + w_{3c} \cdot \texttt{secured} + \epsilon_c,
\vspace{-0.6em}
\end{equation*}

where \(\epsilon_c \sim \mathcal{N}(0, \tau^2)\) introduces stochasticity through a temperature-like parameter. 
The softmax transformation ensures valid probability vectors, from which a categorical label is sampled. 
This design allows us to model recovery-linked variation in categorical variables such as industry sector or region.

\subsubsection{Introducing Controlled Data Shifts}

To investigate how transfer learning methods handle domain changes, we introduce \emph{shifted} variants of our simulation scenario. The idea is to systematically vary key parameters so that the target domain’s data distribution changes while maintaining the same overall structure. In our notation, \(P_S(X)\) and \(P_T(X)\) denote the marginal distributions of the feature space in the source and target domains, respectively. Following the categorization in \citet{Pan2010}, we consider four types of controlled shifts.

First, under \emph{covariate shift}, the feature distributions differ between source and target domains, such that \(P_S(X) \neq P_T(X)\), while the conditional relationship \(P(R \mid X)\) (i.e., the regression coefficients \(\beta_j\)) remains unchanged. Second, under \emph{conditional shift}, the relationship between features and labels changes, so that \(P_S(R \mid X) \neq P_T(R \mid X)\), although the marginal feature distribution \(P(X)\) is held constant. Third, \emph{label shift} refers to changes in the marginal distribution of recovery rates, \(P_S(R) \neq P_T(R)\), for example, through changes in the mixture proportions of the RR distribution, while feature distributions remain nearly identical. Finally, in many real-world applications, domain differences involve a combination of covariate, conditional, and label shifts rather than any single shift type in isolation.

To operationalize these scenarios, we generate paired source and target datasets by modifying the parameters of the data-generating process that governs RR and conditional feature mechanisms. Starting from a shared base configuration (covering sample size, mixture components, and functional forms for \(X_j\)), domain shifts are introduced through structured parameter modifications that target specific distributional components. Covariate shifts are implemented by altering the parameters of the marginal feature-generation functions, such as their means or variances, while keeping the conditional relationship \(P(R \mid X)\) fixed. Conditional shifts modify the parameters linking \(R\) and \(X\), such as regression coefficients or nonlinear transformations, without altering the marginal feature distributions. We induce label shifts by adjusting the parameters of the RR mixture model, particularly the mixture proportions that govern the prevalence of low- and high-recovery regimes. Combined shifts apply multiple perturbations simultaneously. 
These controlled shift %constructions
form the building blocks of the Monte Carlo design described next, where shift type, intensity, schema relation, and sample size are varied jointly.

%---------------------

\subsection{Monte Carlo Simulation Design}
\label{subsec:mc_design}
%Building on the simulation mechanisms above, we develop 
We next present our Monte Carlo framework to study distributional drift, schema mismatch, and limited target data under fully controlled conditions.
This constitutes the fifth step of our evaluation and provides a synthetic environment in which individual factors can be varied systematically, without the confounding effects present in real portfolios.

The simulation generates paired source--target datasets by specifying shift type, shift intensity, source/target schema relation, and target sample size. We consider three types of drift commonly discussed in the TL literature.  
Under \emph{covariate shift}, only the marginal distribution of the features is perturbed. For instance, through location or scale changes or mild nonlinear warping, while the conditional mechanism \(P(R\mid X)\) remains stable.  
\emph{Conditional shift} instead modifies the data-generating relationship between features \(X\) and labels \(R\), keeping the feature marginals \(P(X)\) intact.  
Finally, \emph{label shift} alters the mixture structure of the RR distribution, such as by changing component means or mixture weights, with \(P(X)\) held approximately constant.  
Each of these settings can be combined with different drift intensities, allowing us to trace performance as shifts become more intense.

A second dimension of the simulation concerns the relation between the source and target feature sets.  
We study cases in which the feature schemas are identical (\(\mathcal{F}_{\mathrm{s}} = \mathcal{F}_{\mathrm{t}}\)) as well as asymmetric settings where one domain contains additional features (\(\mathcal{F}_{\mathrm{s}} \neq \mathcal{F}_{\mathrm{t}}\)), including both \(\mathcal{F}_{\mathrm{s}} \subset \mathcal{F}_{\mathrm{t}}\) and \(\mathcal{F}_{\mathrm{t}} \subset \mathcal{F}_{\mathrm{s}}\)).
When schemas differ, we ensure that at least one numeric and one categorical block remain shared and that overall overlap does not fall below 30\%, to create realistic yet manageable levels of heterogeneity.  
To probe data efficiency, target sample sizes range from very small (\(n_t = 100\)) to more substantial (\(n_t = 1000\)).

\begin{figure}[ht!]
\centering
\scalebox{0.92}{
\begin{tikzpicture}[
    node distance=0.5cm and 0.5cm, auto, font=\footnotesize,
    classBlock/.style={rectangle, draw, fill=huberlin-lightblue!60, text width=5em, align=center, rounded corners, minimum height=3em},
    dataBlock/.style={rectangle, draw, fill=huberlin-lightsand, text width=5em, align=center, rounded corners, minimum height=3em},
    targetBlock/.style={rectangle, draw, fill=nus-orange!30, text width=5em, align=center, rounded corners, minimum height=3em},
    arrow/.style={thick, ->, >=stealth}
]
\node[dataBlock] (Config) {Config \\ (type, $s$, schema, $n_t$)};
\coordinate (A) at ($(Config.east)+(0.5,2.6)$);
\coordinate (B) at ($(Config.east)+(12.6,-2.6)$);
\draw[dashed, gray] (A) rectangle (B);
\node at ($(A)!0.5!(B) + (0, 2.65)$) [above, gray, font=\bfseries] {Monte Carlo Simulation};

\node[classBlock, right=1cm of Config] (DataGen) {Data Generation};
\node[dataBlock, above right=0.5cm and 0.5cm of DataGen] (Source) {Source Data};
\node[targetBlock, below right=0.5cm and 0.5cm of DataGen] (Target) {Target Data};
\node[classBlock, right=3cm of DataGen] (Model) {Pretrain \& Fine-tune};
\node[dataBlock, right=1cm of Model] (Eval) {$R^2$, MAE};
\node[classBlock, right=1cm of Eval] (Summary) {Aggregates};

\draw[arrow] (Config) -- (DataGen);
\draw[arrow] (DataGen) -- (Source);
\draw[arrow] (DataGen) -- (Target);
\draw[arrow] (Source) -- (Model);
\draw[arrow] (Target) -- (Model);
\draw[arrow] (Model) -- (Eval);
\draw[arrow] (Eval) -- (Summary);
\end{tikzpicture}
}
\vspace{0.1em}
\caption[
Monte Carlo Simulation Workflow
]{
Monte Carlo simulation workflow from configuration to aggregated results.
}
\label{fig:montecarlo_flow}
\end{figure}
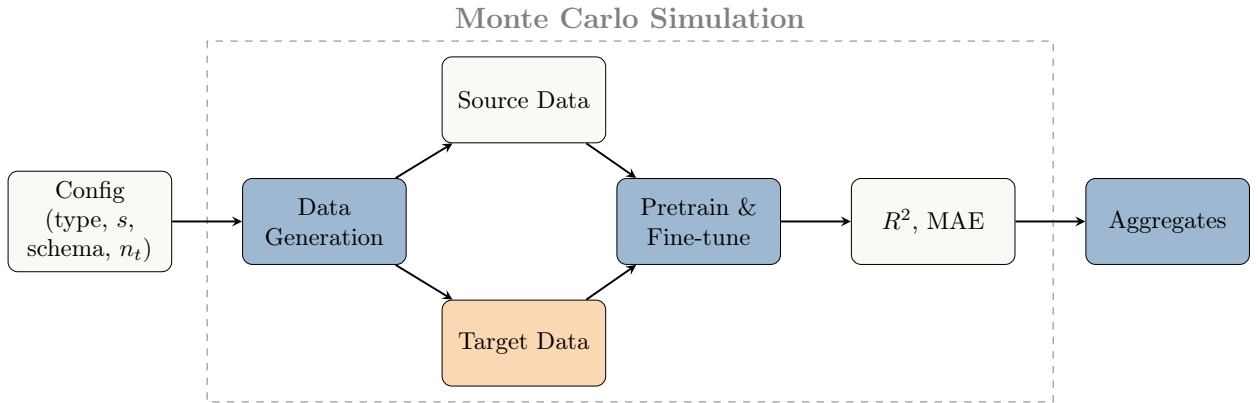

For every combination of shift type, shift intensity, schema relation, and target size, we generate multiple Monte Carlo replications using fixed random seeds, enabling reproducibility and resumable execution.  
Models are pretrained on simulated source data, fine-tuned on the corresponding target data using the same token-level adaptation mechanism as in the real-data setting, and evaluated using the coefficient of determination (primary metric) and MAE (secondary metric).  
Results are summarized using heatmaps, violin plots, and performance curves as a function of the drift parameter.
Figure~\ref{fig:montecarlo_flow} provides an overview of the workflow: after selecting a specific configuration, the simulator produces synthetic source and target datasets with the desired shift pattern. The model is then trained, transferred, and assessed using the metrics above.

\vspace{0.5em}
\subsection{Shift Diagnostics}
\label{subsec:shift_diagnostics}

To characterize each simulated source--target pair, we compute simple empirical drift scores that quantify the effective magnitude of the generated shifts.
For each feature \(X_j\), source and target marginals are approximated by equal-width histograms (with small \(\epsilon\) smoothing), and we compute the directed Kullback--Leibler (KL) divergence from source to target:
\vspace{-0.6em}
\[
D_{\mathrm{KL}}^{\mathrm{s}\to\mathrm{t}}(X_j)
= D_{\mathrm{KL}}\!\big(P_{\mathrm{s}}(X_j)\,\|\,P_{\mathrm{t}}(X_j)\big).
\vspace{-0.6em}
\]
We aggregate per-feature divergences with relevance weights \(w_j=\lvert\mathrm{corr}_{\mathrm{s}}(X_j,R)\rvert\) computed on the source:
\[
\mathrm{FeatureShift}
= \frac{\sum_j w_j\, D_{\mathrm{KL}}^{\mathrm{s}\to\mathrm{t}}(X_j)}{\sum_j w_j}.
\]
Label drift is summarized analogously by the directed divergence between empirical RR histograms,
\[
\mathrm{LabelShift}
= D_{\mathrm{KL}}\!\big(P_{\mathrm{s}}(R)\,\|\,P_{\mathrm{t}}(R)\big).
\]

These diagnostics serve as descriptive indicators of how strongly the target domain departs from the source and help contextualize performance differences observed in the controlled ablations.  
They are not optimization targets themselves but quantify the realized degree of drift resulting from the simulated shift type, intensity, schema relation, and target size.

%-------------

\subsection{Step 5: Simulation Results Under Controlled Shifts}
\label{subsec:sim_results}

Our Monte Carlo experiments systematically evaluate how FT--MDN--T and baseline models respond to different types of distributional drift and schema mismatch, complementing the real-data results by isolating individual transfer mechanisms under controlled simulation settings.
Each simulation draws synthetic datasets with controlled feature distributions, allowing us to vary shift types (covariate, conditional, label), feature-overlap regimes between source and target domains (equal, source~$\subset$~target, target~$\subset$~source), and target sample size. 
Across experiments, we consider FT--MDN--T and its regression-head variant FT-Reg-T, and include XGBoost and an MLP as representative tabular baselines.
To organize the results, we present three complementary perspectives below. 
All perspectives are based on the same underlying Monte Carlo runs but emphasize different aspects of robustness, allowing us to disentangle the effects of feature heterogeneity, shift severity, and data availability.

\begin{figure}[ht!]
    \centering
    \includegraphics[width=0.95\linewidth]{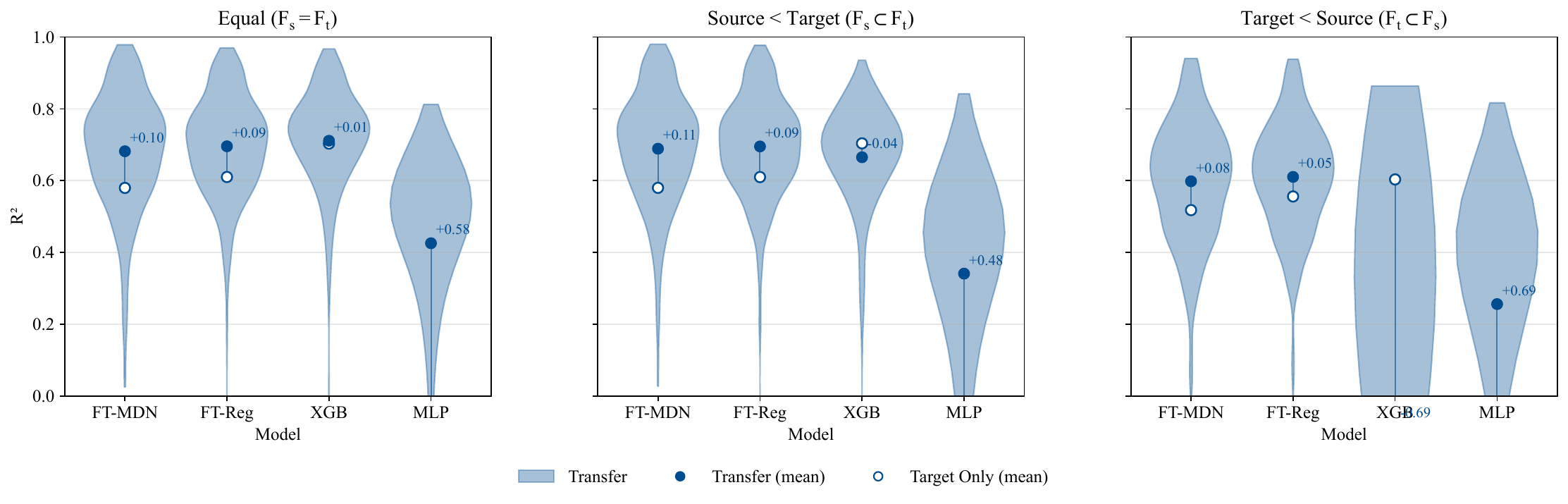}
    \caption[
Transfer vs. Target-Only Performance
]{
Transfer versus target-only performance across heterogeneity regimes.
FT--MDN--T outperforms even with missing features, while XGBoost often fails
under schema mismatch.
}
    \label{fig:violin}
\end{figure}

\paragraph{Robustness across heterogeneity}
We first isolate the effect of feature-space overlap on transfer performance. Using the Monte Carlo runs described above, we group results by overlap regime---equal features (\(F_s = F_t\)), source~$\subset$~target, and target~$\subset$~source---and compare TL to target-baseline training. Figure~\ref{fig:violin} summarizes the resulting validation \(R^2\) values across Monte Carlo runs using violin plots: the vertical extent shows the range of outcomes across runs, while the width indicates where results concentrate (higher density). For each model and regime, the filled marker indicates the mean \(R^2\) under TL, the open marker the corresponding target-baseline mean, and the connecting segment (with the numeric annotation) shows the mean difference between both settings.

Across all overlap regimes, FT--MDN--T achieves consistently higher \(R^2\) than training on the target data alone, with stable gains of roughly \(+0.08\) to \(+0.11\) in mean performance depending on the overlap setting. 
FT-Reg-T shows a very similar pattern, though with slightly smaller gains (around \(+0.05\) to \(+0.09\)). Note that a small number of favorable runs does not drive the gains. They persist across the distribution of Monte Carlo realizations.
We take the observed results as evidence that the Transformer backbone reliably benefits from pretraining even when the feature schema changes between source and target. 

In contrast, XGBoost is considerably more sensitive to schema mismatch. 
While it is largely unaffected when feature sets match (only a marginal gain), its mean performance deteriorates once schemas differ, most notably in the \emph{target $\subset$ source} setting, where features used during pretraining are absent at fine-tuning time, and transfer can collapse (strongly negative mean difference). 
The MLP remains technically compatible across regimes and can achieve significant improvements through transfer. Still, its distributions are wide, and its mean performance remains below the FT-based variants, reflecting higher variability and less reliable generalization. These results show that feature heterogeneity alone does not prevent effective transfer. Performance degrades mainly for models that are sensitive to features appearing or disappearing across domains, whereas FT--MDN--T remains stable across all overlap regimes.

\begin{figure}[ht!]
    \centering
    \includegraphics[width=\linewidth]{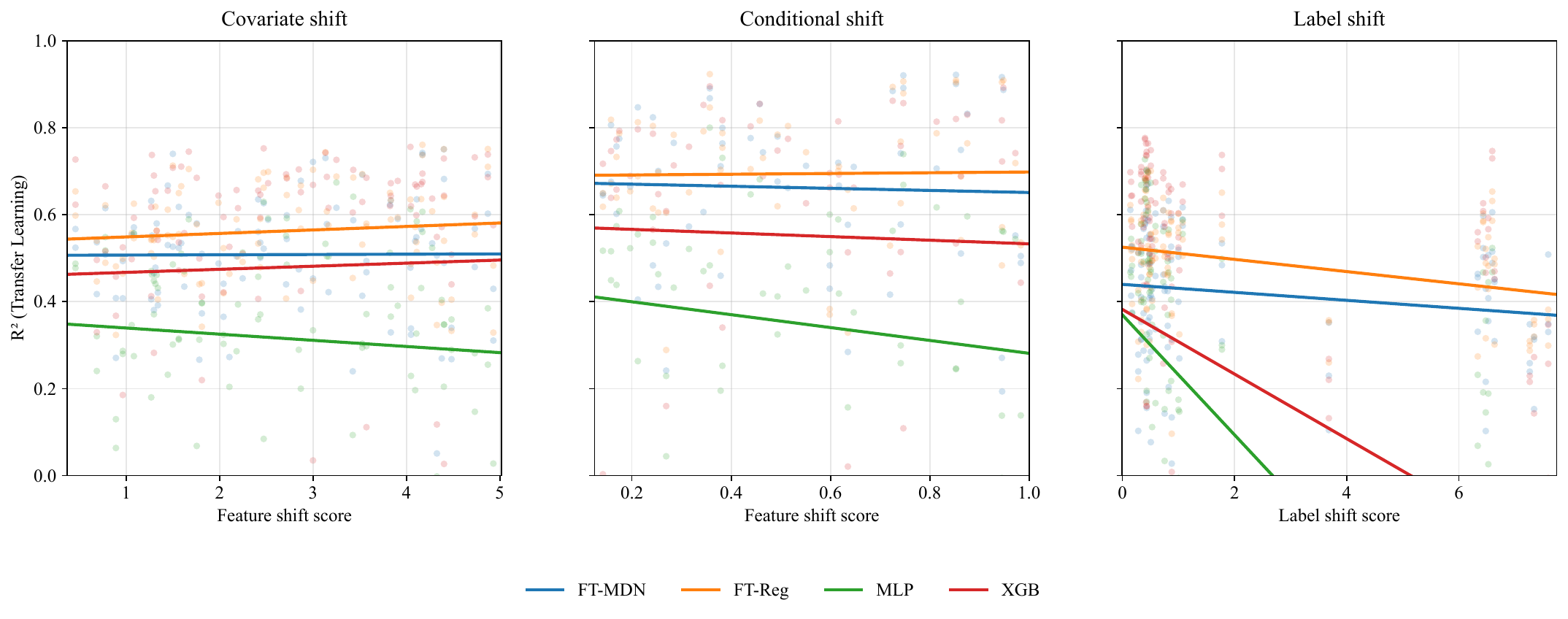}
    \caption[
$R^2$ Degradation Under Shift Intensity
]{
$R^2$ degradation with increasing shift intensity at 100 target samples.
FT--MDN--T is robust under covariate and conditional shifts but deteriorates
under label shift, where tree- and MLP-based models fail sharply.
}
    \label{fig:r2_shift}
\end{figure}

\paragraph{Performance degradation under shift intensity}
Next, we focus on how performance changes as the severity of drift increases. Figure~\ref{fig:r2_shift} plots validation \(R^2\) against a continuous shift score (distributional distance between source and target) for a fixed target size of 100, separately for covariate, conditional, and label shift. For each Monte Carlo run, a new target dataset is generated, and its distributional distance to the source domain is quantified as the shift score on the x-axis. 
We then pretrain each model on the source data and fine-tune it on the corresponding shifted target sample, recording the resulting \(R^2\). 
Each point in the figure represents one such run, and the regression lines depict average performance trends across different shift intensities.
\vspace{-1em}
\vspace{-1em}

Under covariate and conditional shift, all models remain relatively stable, showing minor performance changes, even at higher shift levels. The shift-score scales differ across panels, so slopes should be interpreted within each shift type rather than compared across shift types. FT--MDN--T consistently achieves the highest \(R^2\), followed by FT-Reg-T and XGBoost, while MLP performs worse. %lower overall. 
In these two settings, the fitted trend lines are almost flat, indicating that moderate-to-strong changes in the marginal feature distribution (covariate shift) or in the conditional mapping (conditional shift) do not systematically break transfer performance in our simulation setup. However, there is visible run-to-run variability around the mean trends.

Under label shift, however, performance decreases much more visibly as the label-shift score increases. 
FT--MDN--T and FT-Reg-T decline gradually and remain functional across most of the range, whereas MLP and XGBoost drop off sharply and can approach near-zero \(R^2\) at larger label divergences. This pronounced sensitivity is consistent with the fact that label shift directly alters the RR distribution, so pretrained patterns become less informative even after fine-tuning on \(n_t=100\) target observations. 
These results show that FT--MDN--T is highly resilient to covariate and conditional drift, while substantial label shift remains the dominant challenge for effective transfer.

\begin{figure}[ht!]
    \centering
    \includegraphics[width=\linewidth]{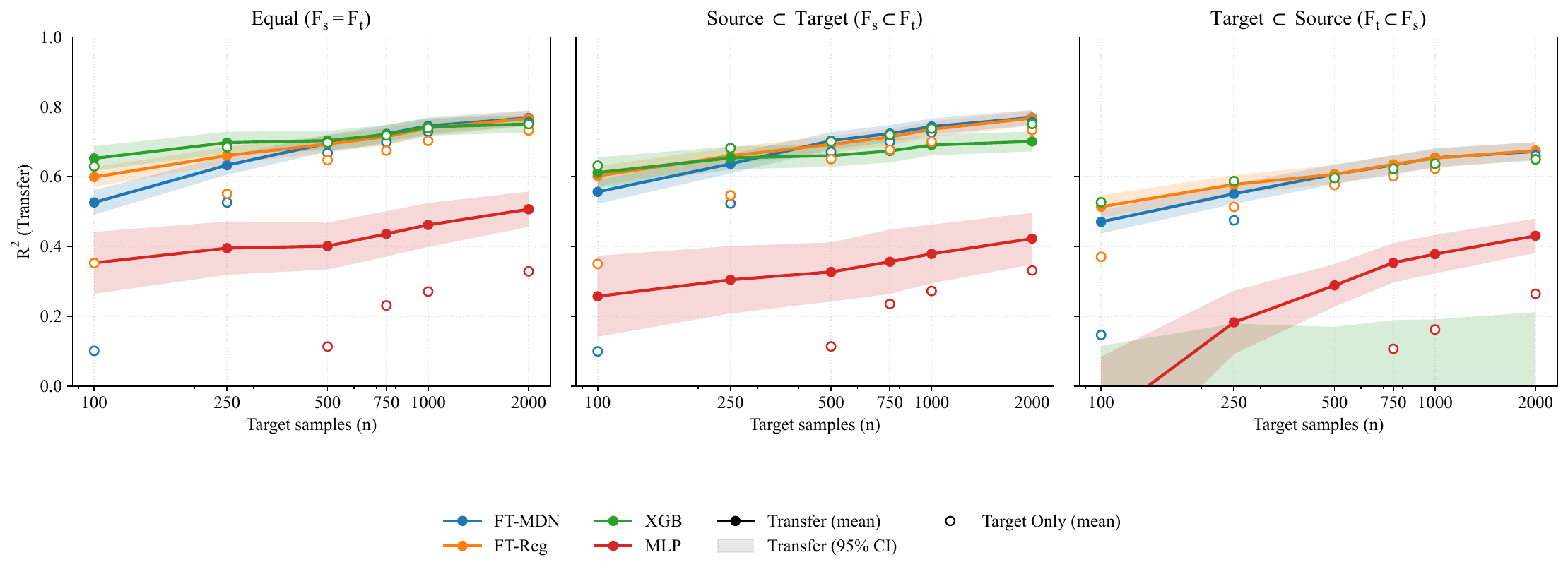}
    \caption[
Effect of Target Sample Size
]{
Effect of target sample size across heterogeneity regimes.
Transfer accelerates learning, particularly when target data are scarce.
}
    \label{fig:sample_size}
\end{figure}

\paragraph{Sample efficiency}
Finally, we examine how the benefit of TL depends on the amount of available target data. 
For each heterogeneity regime---equal features, source~$\subset$~target, and target~$\subset$~source---we vary the target sample size \(n_t\) and evaluate validation \(R^2\) after fine-tuning. 
Each point in Figure~\ref{fig:sample_size} represents the mean performance across multiple Monte Carlo runs, with shaded areas denoting 95\% confidence intervals.

FT--MDN--T with transfer consistently accelerates learning: at small sample sizes (e.g., \(n_t = 100\)), the transfer markers (filled points) lie clearly above the corresponding target-only markers (open points), while the gap narrows as more target data become available. This pattern holds across all three overlap regimes, indicating that pretraining provides a systematic advantage in the small-data regime.

Comparing models, however, the picture depends strongly on whether feature spaces match. If they are identical (\(F_s = F_t\), left panel), XGBoost remains highly competitive and can outperform the FT-based variants at smaller \(n_t\). Once feature heterogeneity is introduced (middle and right panels), the ranking shifts: FT--MDN--T and FT-Reg-T maintain strong performance, whereas XGBoost becomes less reliable. %and no longer dominates.
This difference is most pronounced in the \emph{target $\subset$ source} setting, where XGBoost transfer can degrade substantially, while the FT-based variants remain stable across target sizes. The MLP model follows the same qualitative trend but remains consistently below the FT-based models across regimes and target sizes, with no clear advantage in either homogeneous or heterogeneous settings.

The confidence bands shrink as \(n_t\) increases, reflecting reduced uncertainty with more target data. When target data are extremely limited, the regression-head variant (FT-Reg-T) sometimes achieves slightly higher \(R^2\) than the MDN version, reflecting greater stability in very small-sample regimes. As the sample size grows, however, FT--MDN--T ultimately reaches the best overall performance. While differences in \(R^2\) between FT-Reg-T and FT--MDN--T become smaller in this regime, the MDN-based variant retains the advantage of modeling the full conditional RR distribution, which is particularly relevant for bimodal recovery outcomes.

\subsection{Synthesis and Implications}
\label{subsec:synthesis}

For completeness, we provide FT--MDN--T performance across the complete simulation grid (shift type $\times$ overlap regime $\times$ shift intensity $\times$ target size) in the Online Appendix. Across this grid, a consistent pattern emerges: covariate and conditional shifts have a comparatively mild effect across overlap regimes, whereas label shift drives the strongest, most systematic performance degradation. Increasing the number of target samples mitigates these effects, but strong label divergence remains the most difficult setting. This broad pattern complements the real-data GCD~$\rightarrow$~UP5 findings, in which domain mismatch and label distribution differences limit transfer gains even when substantial pretraining information is available.

Bringing together the empirical and simulation results, several overarching insights emerge. 
First, FT--MDN--T’s ability to predict full conditional distributions is a substantive modeling advantage: it captures multimodal recovery behavior, mitigates mean collapse, and achieves closer alignment with empirical recovery patterns---even in cases where traditional point metrics appear satisfactory. Such alignment is crucial for RR modeling, where bimodality is a defining characteristic at both the loan and portfolio levels, and where risk decisions often depend on the shape of the loss distribution rather than a single expected value.

Second, transfer across heterogeneous feature spaces is both feasible and effective, but the details of schema mismatch matter. 
The embedding- and token-masking mechanism enables the model to accommodate missing or newly introduced features, allowing pretrained representations to adapt smoothly instead of breaking when schemas differ. This flexibility is reflected across our experiments: categorical embeddings provide a natural bridge across portfolios with differing category sets, the controlled within-domain schema expansion confirms that the model can incorporate new target-only features without relearning from scratch, and the cross-portfolio results show that pretraining on a cleanly aligned shared subset can be more effective than pretraining on a richer but partially incompatible source schema. In the simulation study, this translates into stable transfer gains across overlap regimes, whereas schema-dependent baselines can degrade sharply when features appear or disappear between pretraining and fine-tuning.

Third, distributional shifts matter, but their impact is not uniform. In our simulations, covariate and conditional drift have a comparatively limited effect across overlap regimes, whereas label shift is consistently the most challenging case. When the RR distribution itself changes substantially, the benefit of pretraining diminishes even after fine-tuning, which helps explain why cross-portfolio transfer from loans to bonds yields only modest improvements. At the same time, the experiments indicate that FT-based models can be sample-efficient in target-only baseline settings. They also evidence that transfer provides its largest benefits primarily in the most data-scarce regimes and under moderate label divergence.

From a practical perspective, our findings indicate that TL offers the largest benefits when (i) target recovery datasets are small, yet large enough to enable effective fine-tuning, (ii) there exists at least moderate feature overlap with historical or related portfolios, and (iii) label distributions are sufficiently aligned. In operational environments, continuous monitoring of domain drift---using indicators such as the KL divergence---is therefore critical, not only for detecting feature drift but also for flagging shifts in the recovery distribution that can directly undermine transfer. Moreover, model choice should reflect the expected degree of schema stability: tabular baselines remain competitive when feature spaces are identical, whereas the schema-flexible FT variants provide more reliable transfer behavior under feature heterogeneity.

\section{Discussion and Conclusion}
\label{sec:conclusion-rr}

In conclusion, this study clarifies when and how TL can improve RR forecasting in small-data regimes, particularly under heterogeneous feature spaces and distributional shifts. We proposed the \textit{FT--MDN--Transformer} (FT--MDN--T)---a tabular Transformer architecture with a mixture-density output head---that accommodates non-identical feature sets across source and target domains (via PAD tokens and attention masking), produces full conditional density forecasts, and is shown to perform robustly under covariate and conditional shift, while degrading under substantial label shift.

\medskip

\noindent\textbf{Main findings.}  
From the combined empirical and simulation evidence, three central conclusions emerge. First, RR benefits from distributional modeling: FT--MDN--T captures multimodal recovery behavior and provides portfolio-level density forecasts that preserve risk-relevant structure, which point metrics alone can obscure. Second, TL across heterogeneous feature spaces is feasible and practically useful: across real-data and simulation settings, FT--MDN--T remains stable when features appear or disappear between pretraining and fine-tuning, whereas schema-dependent baselines can degrade sharply under mismatch. Third, TL primarily acts as a sample-efficiency mechanism: it yields the largest gains when target recovery data are limited. At the same time, the relative advantage diminishes as more target observations become available and target-only training can catch up.

\medskip

\noindent\textbf{Interpretation and contributions.}  
These findings show that combining token-level embeddings with mixture-density outputs creates a flexible and effective architecture for credit recovery modeling. The PAD-token and masking mechanism enable smooth adaptation across heterogeneous feature spaces without manual alignment, allowing pretrained models to remain functional even when the target schema differs. At the same time, the results indicate a clear boundary condition for transfer. When the recovery distribution itself shifts substantially (label shift), the benefit of pretraining diminishes even after fine-tuning. This implies that, in real-world portfolio transfers---especially across asset classes or collateral structures---practitioners should explicitly assess marginal recovery-distribution drift and consider corrective measures before deploying transfer models. More broadly, the study provides conceptual and empirical evidence that distributional modeling and schema-flexible architectures can make transfer learning viable in complex, heterogeneous financial domains.

From a practitioner’s perspective, TL is most promising when target recovery data are limited but still sufficient to support fine-tuning, feature overlap with historical datasets is at least moderate, and RR distributions between source and target remain broadly aligned. In operational use, continuous monitoring of domain drift---through measures such as the KL divergence---is essential to identify when retraining or adaptation becomes necessary, and to flag recovery-distribution shifts that can directly undermine transfer.

\medskip

\noindent\textbf{Limitations and future directions.}  
While this work advances the understanding of TL for RR prediction, several limitations remain. The current FT--MDN--T architecture does not include explicit label-shift correction, even though this is the scenario where performance degrades most strongly. Our simulation framework is based on mixture-Gaussian assumptions and simplified functional relationships, which may not capture the full complexity of real-world recovery processes. Moreover, the study focuses on supervised settings; extending FT--MDN--T to semi-supervised or domain-adversarial regimes, and integrating explicit label-shift correction are promising directions to enhance robustness in more challenging transfer scenarios further. Finally, since the MDN head provides full conditional densities that are operationally meaningful for bimodal RR, future work could place more emphasis on distribution-sensitive evaluation and calibration analyses alongside point metrics to better reflect downstream use cases such as provisioning, stress testing, and tail-risk assessment.

\medskip

In summary, this study shows that TL can meaningfully improve RR forecasting in small-data regimes when supported by architectures that accommodate schema heterogeneity and model full conditional distributions. Nonetheless, label shift remains the central barrier to effective cross-domain transfer. We expect that the proposed architecture, empirical findings, and simulation framework will provide a foundation and practical guide for future research and deployment of TL in credit-risk and related financial modeling applications.

\appendix

\appendix
\label{appendix-rr}

\section{Additional Implementation Details}
\label{app:impl}

This appendix summarizes the model configurations used for baselines and ablations. 
Additional implementation details, extended results, and simulation diagnostics are provided in the Online Appendix.
Architectural details of FT--Transformer are given in Sec.~\ref{sec:methodology-rr}. 
Complete code, hyperparameters, and reproducibility scripts are available in the online supplementary material and the accompanying repository.\footnote{\url{https://github.com/christopher3996/recovery_rates_transfer_learning}}

\subsection{Baseline models and ablation settings}

Table~\ref{tab:candidate_models} lists the baseline models, their architectural specifications, and how they participate in the transfer-learning setup.  
Unless otherwise noted, all models share identical data splits, preprocessing, and early-stopping logic, as detailed in the GitHub repository.

\begin{table}[ht!]
\centering
\caption{Baseline models with architectural specifications and transfer mechanisms.}
\label{tab:candidate_models}
\small
\renewcommand{\arraystretch}{1.2}
\begin{tabular}{l p{6cm} p{6cm}}
\toprule
\textbf{Model} & \textbf{Architecture} & \textbf{Transfer Mechanism} \\
\midrule
MLP & Two hidden layers (GELU), tuned learning rate and batch size & Pretrain on shared source features; fine-tune on full target schema \\
XGBoost & Gradient-boosted trees with tuned depth/regularization & Sequential pretraining on shared features; extend to target-only features \\
%Random Forest & Tuned number of estimators and depth & Sequential training on shared features only \\
\midrule
FT--Reg (ablation) & FT--Transformer backbone with regression head & Same schema-masked transfer protocol as main model \\
FT--MDN--Transformer & FT--Transformer backbone + MDN head & Two-stage pretrain$\rightarrow$fine-tune with schema masking \\
\bottomrule
\end{tabular}
\end{table}

\subsection{Hyperparameter summary}

Neural models (MLP, FT--Reg, FT--MDN--T) are trained with AdamW, early stopping, and standard batch sizes.  
Exact hyperparameters, configuration files, and scripts for reproducing all experiments are provided in the online repository.

%\iffalse

%\fi

\newpage

% \bibliographystyle{elsarticle-num-names} 
  \bibliographystyle{elsarticle-harv} 
 \bibliography{ref-recovery-rates}





% \end{thebibliography}
\end{document}